\documentclass[12pt,preprint]{aastex}

\newcommand{\msun}{M_{\odot}}
\newcommand{\be}{\begin{equation}}
\newcommand{\ee}{\end{equation}}

\shorttitle{SNe in CS Bubbles}
\shortauthors{Dwarkadas.}

\begin{document}

%% LaTeX will automatically break titles if they run longer than
%% one line. However, you may use \\ to force a line break if
%% you desire.

\title{The Evolution of Supernovae in Circumstellar Wind-Blown Bubbles
I. Introduction and One-Dimensional Calculations.}

%% Use \author, \affil, and the \and command to format
%% author and affiliation information.
%% Note that \email has replaced the old \authoremail command
%% from AASTeX v4.0. You can use \email to mark an email address
%% anywhere in the paper, not just in the front matter.
%% As in the title, use \\ to force line breaks.

\author{Vikram V. Dwarkadas} 

\affil{Astronomy and Astrophysics, Univ
of Chicago, 5640 S Ellis Ave RI 451, Chicago IL 60637 \\ \ \\ {\bf Submitted to the Astrophysical Journal}}

\email{vikram@oddjob.uchicago.edu}

%% Notice that each of these authors has alternate affiliations, which
%% are identified by the \altaffilmark after each name.  Specify alternate
%% affiliation information with \altaffiltext, with one command per each
%% affiliation.
%% Mark off your abstract in the ``abstract'' environment. In the manuscript
%% style, abstract will output a Received/Accepted line after the
%% title and affiliation information. No date will appear since the author
%% does not have this information. The dates will be filled in by the
%% editorial office after submission.

\begin{abstract}
Mass loss from massive stars ($\ga 8 \msun$) can result in the
formation of circumstellar wind blown cavities surrounding the star,
bordered by a thin, dense, cold shell. When the star explodes as a
core-collapse supernova (SN), the resulting shock wave will interact
with this modified medium around the star, rather than the
interstellar medium. In this work we first explore the nature of the
circumstellar medium around massive stars in various evolutionary
stages. This is followed by a study of the evolution of SNe within
these wind-blown bubbles. The evolution depends primarily on a single
parameter $\Lambda$, the ratio of the mass of the dense shell to that
of the ejected material. We investigate the evolution for different
values of this parameter. We also plot approximate X-ray surface
brightness plots from the simulations. For very small values $\Lambda
<< 1$ the effect of the shell is negligible, as one would
expect. Values of $\Lambda \la 1$ affect the SN evolution, but the SN
'forgets' about the existence of the shell in about 10 doubling times
or so. The remnant density profile changes, and consequently the X-ray
emission from the remnant will also change. The initial X-ray
luminosity of the remnant is quite low, but interaction of the shock
wave with the dense circumstellar shell can increase the luminosity by
2-3 orders of magnitude.  As the reflected shock begins to move
inwards, X-ray images will show the presence of a double-shelled
structure. Larger values result in more SN energy being expended to
the shell. The resulting reflected shock moves quickly back to the
origin, and the ejecta are thermalized rapidly. The evolution of the
remnant is speeded up, and the entire remnant may appear bright in
X-rays. If $\Lambda >> 1$ then a substantial amount of energy may be
expended in the shell. In the extreme case the SN may go directly from
the free-expansion to the adiabatic stage, bypassing the Sedov
stage. Our results show that in many cases the SN remnant spends a
significant amount of time within the bubble. The low density within
the bubble can delay the onset of the Sedov stage, and may end up
reducing the amount of time spent in the Sedov stage. The complicated
density profile within the bubble makes it difficult to infer the
mass-loss properties of the pre-SN star by studying the evolution of
the resulting supernova remnant.

\end{abstract}

%% Keywords should appear after the \end{abstract} command. The uncommented
%% example has been keyed in ApJ style. See the instructions to authors
%% for the journal to which you are submitting your paper to determine
%% what keyword punctuation is appropriate.

%% Authors who wish to have the most important objects in their paper
%% linked in the electronic edition to a data center may do so in the
%% subject header.  Objects should be in the appropriate "individual"
%% headers (e.g. quasars: individual, stars: individual, etc.) with the
%% additional provision that the total number of headers, including each
%% individual object, not exceed six.  The \objectname{} macro, and its
%% alias \object{}, is used to mark each object.  The macro takes the object
%% name as its primary argument.  This name will appear in the paper
%% and serve as the link's anchor in the electronic edition if the name
%% is recognized by the data centers.  The macro also takes an optional
%% argument in parentheses in cases where the data center identification
%% differs from what is to be printed in the paper.


%% From the front matter, we move on to the body of the paper.
%% In the first two sections, notice the use of the natbib \citep
%% and \citet commands to identify citations.  The citations are
%% tied to the reference list via symbolic KEYs. The KEY corresponds
%% to the KEY in the \bibitem in the reference list below. We have
%% chosen the first three characters of the first author's name plus
%% the last two numeral of the year of publication as our KEY for
%% each reference.

\section{Introduction}

Core collapse supernovae (SNe) arise from the collapse of massive
stars with mass M $\ga 8 \msun$. As these stars evolve along the main
sequence and beyond it, they lose a considerable amount of mass in the
form of winds \citep{kp00}. The collective action of the winds from
different evolutionary stages may sweep up the material in the ambient
medium to form a wind-blown cavity, bordered by a thin, dense, cold
shell of material, which contains most of the mass of the swept-up
medium. When the star explodes as a SN, the shock wave resulting from
the explosion evolves in this highly modified circumstellar medium,
and not in the pristine interstellar medium \citep{m88}. The evolution
of the supernova remnant is therefore very different from the
classical evolution in a constant density interstellar medium, where
the remnant successively advances through the phases of
free-expansion, adiabatic or Sedov stage, radiative stage and
dispersal into the surrounding medium \citep[see for example][]{w72}.

Evidence has built up over the last few years that the medium around a
pre-SN star is considerably modified by the action of winds, radiation
and stellar outbursts. This is highlighted by the famous Hubble Space
Telescope picture of the three-ring nebula surrounding the SN
1987A. The combined action of winds and radiation in sculpting this
medium is undeniable \citep{bl93, cd95}. Other known cases of SNe
exploding in wind-driven shells are less spectacular. There are
indications that the Cygnus Loop is the remnant of an old supernova
explosion that went off in a wind-blown bubble
\citep{l97}. Circumstellar interaction models were earlier proposed
for N49 \citep{sdkw85} and N132D \citep{h87}. \citet{bsbs96} have
modeled the X-ray emission from Cas A as arising from a SN explosion
within a wind-blown cavity.

Examples of wind-blown bubbles around massive stars are far more
numerous \citep{c03, cec03}. Ring nebulae are seen around main
sequence O and B stars, OfPe/WN9 stars \citep{n99}, and many
Wolf-Rayet stars in general. Bipolar nebulae are observed around many
LBV stars \citep{w01}, the best-known and studied one being $\eta$
Carinae \citep{dh97}. Circumstellar nebulae are observed around red
supergiant stars such as VY CMa \citep{shd01, hds02}, and around blue
supergiant stars such as NGC6164-5 \citep{lck87}. The progenitor of SN
1987A is thought to be a blue-supergiant (BSG). A similar structure is
seen around the star Sher 25, suggesting that it may be an analogue of
SN 1987A \citep{c03}.

The evolution of remnants within wind-blown cavities has merited some
attention in the past. The problem was discussed numerically by
\citet{cd89}. It was placed on a firm theoretical footing with
analytic calculations by \citet{cl89}. Early numerical calculations
are described in a series of papers by Tenorio-Tagle and
collaborators, wherein they discussed the 1-dimensional evolution
\citep{tbf90}, the 2-dimensional evolution \citep{trfb91}, and
off-center explosions within pre-existent bubbles \citep{rtfb93}. A
review is presented in \citep{ftbr91}. \citet{bl93} and \citet{ma95}
carried out hydrodynamical simulations of the nebula surrounding SN
1987A, and \citet{cd95} incorporated the effects of the ionizing
radiation from the pre-SN star in order to explain the linearly
increasing X-ray and radio emission from SN 1987A, as well as the
considerable slowing down of the shock front. \citet{bsbs96} modeled
the X-ray emission from Cas A under the assumption that the SN
expanded in a wind-blown bubble.

In recent years the availability of high spectral and spatial
resolution data, both space and ground-based, has added considerably
to our knowledge of circumstellar (CS) wind-blown bubbles surrounding
massive stars. Observations from HST, Chandra, XMM, and now SIRTF have
enriched our knowledge of the structure, formation and evolution of
wind-blown nebulae. Simultaneously, advances in understanding the
structure of the ejected material in supernovae \citep[for
e.g.][]{tm99, mm99} have further increased our understanding of
supernova evolution. The worldwide effort gone into studying SN 1987A
has enabled us to explore in real time a SN expanding within a
wind-blown bubble on its way to becoming a SNR.

The recent confirmation that gamma-ray bursts arise from very massive
stars led to the realization that the shock waves in this case may
also evolve in the pre-outburst stellar winds from the progenitor,
rather than in the ISM. There is some speculation that gamma-ray burst
afterglows may be explained in this manner \citep{clf04}. The extreme
interest in gamma-ray bursts as cosmological phenomena has given
further impetus to the study of supernova shock waves in stellar
winds.

Given the considerable increase in our observational knowledge,
accompanied by improved theoretical understanding of SN evolution and
SN progenitors, we wish to revisit the evolution of SNRs in
circumstellar wind-blown bubbles. We have planned a comprehensive
effort to explore the dynamics, kinematics and radiation signatures of
SNe expanding in CS wind-blown bubbles. In this first paper we will
explore the various aspects of the interaction, define the parameters
on which the evolution depends, and carry out a broad parameter
survey. Our initial models will be idealized in order to understand
the effects of various parameters. In subsequent papers we will take
more realistic stellar evolution calculations into account, which
provide the wind-properties from a star at each stage. These will help
us to explore the structure and kinematics of more complicated
objects. Finally we will apply the knowledge gained to investigations
of specific astrophysical objects, especially SN 1987A.

The remainder of this paper is organized as follows. In \S
\ref{sec:massloss} we briefly discuss mass-loss from massive
stars. Hot and cool stars are considered separately.  In \S
\ref{sec:wbb} we describe the basic structure of CS wind-blown
bubbles, and in \S \ref{sec:snejecta} we outline the density structure
of the ejected SN material. \S \ref{sec:ts} describes a simple
analytic solution that helps to understand the basic ideas, and sets
the stage for the numerical calculations. A very brief overview of the
code is given in \S \ref{sec:numinfo}.  In \S \ref{sec:snbub} we study
the basic features of the interaction of SN shock waves with
circumstellar wind-blown bubbles, using spherically symmetric
simulations. Two specific cases are chosen to illustrate the basic
aspects of the interaction. X-ray brightness profiles and luminosity
plots are presented.  \S \ref{sec:disc} discusses the implications of
our research for the onset and growth of the Sedov stage. Finally, \S
\ref{sec:summary} summarizes the basic results, and outlines future
work.

\section{Stellar Mass Loss}
\label{sec:massloss}

Core-Collapse SNe (which includes the classes Type II, as well as Type
Ib/c) arise from the gravitational collapse of massive stars. The line
dividing massive versus non-massive stars is usually drawn at about 8
$\msun$, which is the mass at which a star will not eventually turn
into a white dwarf \citep{l94}. This however still leaves a large
range of stars, with initial main sequence mass between 8 $\sim$ 200
$\msun$, that can explode as SNe. The evolution and mass-loss
characteristics of these stars are quite different over this large
range. Thus the environment into which a core-collapse SN evolves will
be considerably different depending on the initial mass of the star.

The velocity of the wind emitted from a star is generally comparable
to or a few times larger than the escape velocity of the star, as
would be expected. The mass loss rate depends on what drives the
material from the stellar surface \citep{lc99}. Stars like the sun
have coronal, pressure driven winds with a very low mass-loss
rate. Hot, luminous stars such as O, B and Wolf-Rayet (WR) stars have
radiatively driven winds. The mass-loss mechanism in cool evolved
stars, such as red supergiants (RSGs), is not well understood,
although it is suspected to be due to radiation pressure on dust
grains.

The two parameters, mass-loss rate and wind-velocity, determine the
density of the wind around the star. For a wind with constant
properties, the density $\rho$ at a radius $r$ will be given by $\rho
= \dot{M}/(4\pi r^2 v_w)$ where $\dot{M}$ is the mass-loss rate and
$v_w$ is the wind velocity. In general we can differentiate two clear
regimes:

\subsection{Mass Loss from cool massive stars:} 

Most stars larger than about 11 $\msun$ will become red supergiant
stars \citep{ah04}. The lack of RSGs above 50-60 $\msun$ appeared to
indicate the existence of an upper limit to the RSG stage. Conflicting
reports about this result have been recently clarified by
\citet{lnpsl01}.  More importantly for this paper, most single stars
below about 35 $\msun$ will explode as SNe while in the RSG stage
\citep{ah04}. These results are quoted only for solar metallicities,
for lower metallicities the number of RSGs reduces considerably. Plus
rotational mixing, semi-convection and overshoot will change these
numbers. Nevertheless, since the number of stars decreases sharply
with increasing mass, the result generally implies that a large
fraction of core-collapse SNe arise from RSG stars.

While the mass loss process in red supergiants is not generally well
understood, they are observed to have very slow winds, on the order of
20-100 km/s, with high mass loss rates on the order of a few times
10$^{-5}$ to 10$^{-4}$ $\msun$/yr. The mass loss is not uniform, but
could vary over time periods of a few thousand years. One of the best
observed post-RSG stars, IRC +10420, shows evidence for an
increasingly complex environment around the star, with multiple
shells, several individual condensations and a generally asymmetric
structure \citep{hsd97}. These authors infer a mass-loss rate about
10$^{-6} \msun$/yr in the main-sequence (MS) phase, increasing to few
times 10$^{-4}$ to 10$^{-3} \msun$/yr during various stages of
evolution. Another extreme RSG star VY CMa also shows evidence of an
asymmetric structure, reflection arcs and knots, and several bright
condensations \citep{shd01}. The deduced mass-loss rate is about 3
$\times 10^{-4} \msun$/yr.

The high mass loss rate and low wind-velocity imply a high density for
the circumstellar material around the star. In general the density can
be written as:

\begin{equation}
{\rho_{RSG}} \sim 5 \times 10^{-20}~{\dot{M}_{-4}}~ {r_{17}^{-2}}~ v_{1}^{-1}
\end{equation}

\noindent
where ${\dot{M}_{-4}}$ is the mass-loss rate in units of 10$^{-4}
\msun$, {r$_{17}$} is the radius in units of 10$^{17}$ cms and $v_{1}$
is the wind-velocity in units of 10 km/s. Assuming that the RSG stage
lasts for 10$^5$ years, this RSG wind can alter the medium to a
distance of a few parsecs.

\subsection{Mass Loss from Hot Massive Stars}

Stars above about 35 $\msun$ suffer from considerable mass-loss which
tends to strip them of their H envelopes. These Wolf-Rayet (WR) stars
are presumed to be the progenitors of Type Ib/c SNe, which do not show
H lines in their spectrum. These stars are in general hot, massive
stars, which lose mass via radiatively driven winds from the stellar
surface. Their progenitors are mainly early type O and B stars, with
mass-loss rates on the order of 10$^{-8}$ to 10$^{-5} \msun$/yr, and
terminal wind velocities about 1000-3000 km/s \citep{djnv88}. When the
star evolves off the main sequence stage and into the WR phase (with
perhaps some intermediary stages depending on the initial mass), the
mass loss rate can increase to a few times 10$^{-5} \msun$/yr
\citep{k97}. Thus these stars can lose a large fraction of their
stellar mass during their lifetime. For example, in one of Norbert
Langer's models, a 35 $\msun$ star has only about 9 $\msun$ remaining
when it explodes as a supernova, with 26 $\msun$ being deposited in
the surrounding medium. This material can form large wind-blown
cavities stretching for tens of parsecs. The various evolutionary
stages may lead to the formation of multiple wind-blown shells, which
may not always be visible.

Given the large wind-velocities of MS and W-R hot stars, the density
of the circumstellar medium will be considerably reduced as compared
to the RSG stars mentioned above. The density from a W-R wind at a
radius $r$ from the star is given by

\begin{equation}
{\rho_{WR}} \sim 2.5 \times 10^{-23}~{\dot{M}_{-5}}~ {r_{17}^{-2}}~ v_{3}^{-1}
\end{equation}

\noindent
where ${\dot{M}_{-5}}$ is the mass-loss rate in units of 10$^{-5}
\msun$, {r$_{17}$} is the radius in units of 10$^{17}$ cms and $v_{3}$
is the wind-velocity in units of 10$^3$ km/s.

It is clear that the density of the wind around a WR star is about 2-3
orders of magnitude less than that around a RSG star. This has
important consequences for the SN evolution, since much of the
emission from the young remnant is due to the interaction with the
circumstellar medium \citep{cf94}. In most cases this depends on the
density of the surrounding medium, and will therefore be considerably
reduced in a lower-density medium.

The above discussion shows that while SNe arise from both RSG and WR
progenitors, the immediate medium into which the shock wave evolves
will be quite different. Both of the stages are post-main sequence
stages, and the WR stage may sometimes even be a post-RSG stage. In
general the wind from the star in each case will evolve inside a
previously evacuated wind-blown cavity carved out by the MS stage. An
important distinction however is that the wind from the RSG star is
usually slower than the wind from the preceding stage and will not
shock it, but will result in a new pressure equilibrium. The fast WR
wind will drive a strong shock wave into the surrounding medium. These
variations can lead to very different structures around the star.

What is clear is that in many cases a SN will expand into a
low-density environment during its initial evolution, with density
lower than the surrounding interstellar medium. In the case of SNe
from hot massive stars, this low density medium will start right from
the star itself. In the case of RSG progenitors, there may exist a
very high-density medium around the star, in which the remnant may
evolve for a few tens to hundreds of years. Following this the remnant
will continue to evolve for some time in a low-density medium formed
by the wind from the main-sequence star, usually an O or B star. In
this work therefore, we have concentrated on remnants evolving in a
low-density circumstellar wind-blown bubble.

\subsection{Other Considerations}

The evolution of single, massive stars to the supernova stage has
received considerable attention. However many stars have one or more
companions, the presence of which can considerably alter the evolution
of the star.  Mass transfer from or to a companion star can slow down
or speed up the evolution of the star, alter its evolutionary track,
and even result in a stellar explosion as in the case of Type Ia
supernovae. The best studied supernova, SN 1987A, is known to have a
blue supergiant progenitor, a B3Ia star when it exploded. A blue
supergiant progenitor was unexpected, as it indicated that the star on
the H-R diagram went from the blue to the red and back to the blue
region before it exploded. Although some single-star scenarios have
managed to produce a blue supergiant before the star expolodes
\citep[see for example][]{whwl97}, they require very low
metallicities.  Rotation changes the dynamics somewhat, and may also
help to explain the origin of the bipolar circumstellar shell
surrounding SN 1987A, but still appears to require special
conditions. A BSG progenitor has consequently been suggested as
indicative of binary evolution \citep{p92}. While conclusive evidence
is lacking, it does imply that stars in binary systems evolve
differently from single stars, and SN progenitors in binary systems
may differ from single-star progenitors. Not surprisingly, the medium
around the star will also be correspondingly different. SN 1987A
clearly shows the presence of a beautiful bipolar circumstellar shell
structure, surrounded by a thin dense shell. However the density
within the bubble is neither as large as that in RSGs nor as low as
that in W-R stars, but somewhat intermediate between the two, on the
order of 1 particle/cc. A region of ionized wind material, with a
density of about 100 particles/cc, also exists inner to the
circumstellar nebula \citep{cd95}.

\section{Circumstellar Wind-Blown Bubbles}
\label{sec:wbb}
\begin{figure}
\includegraphics[angle=90,scale=0.85]{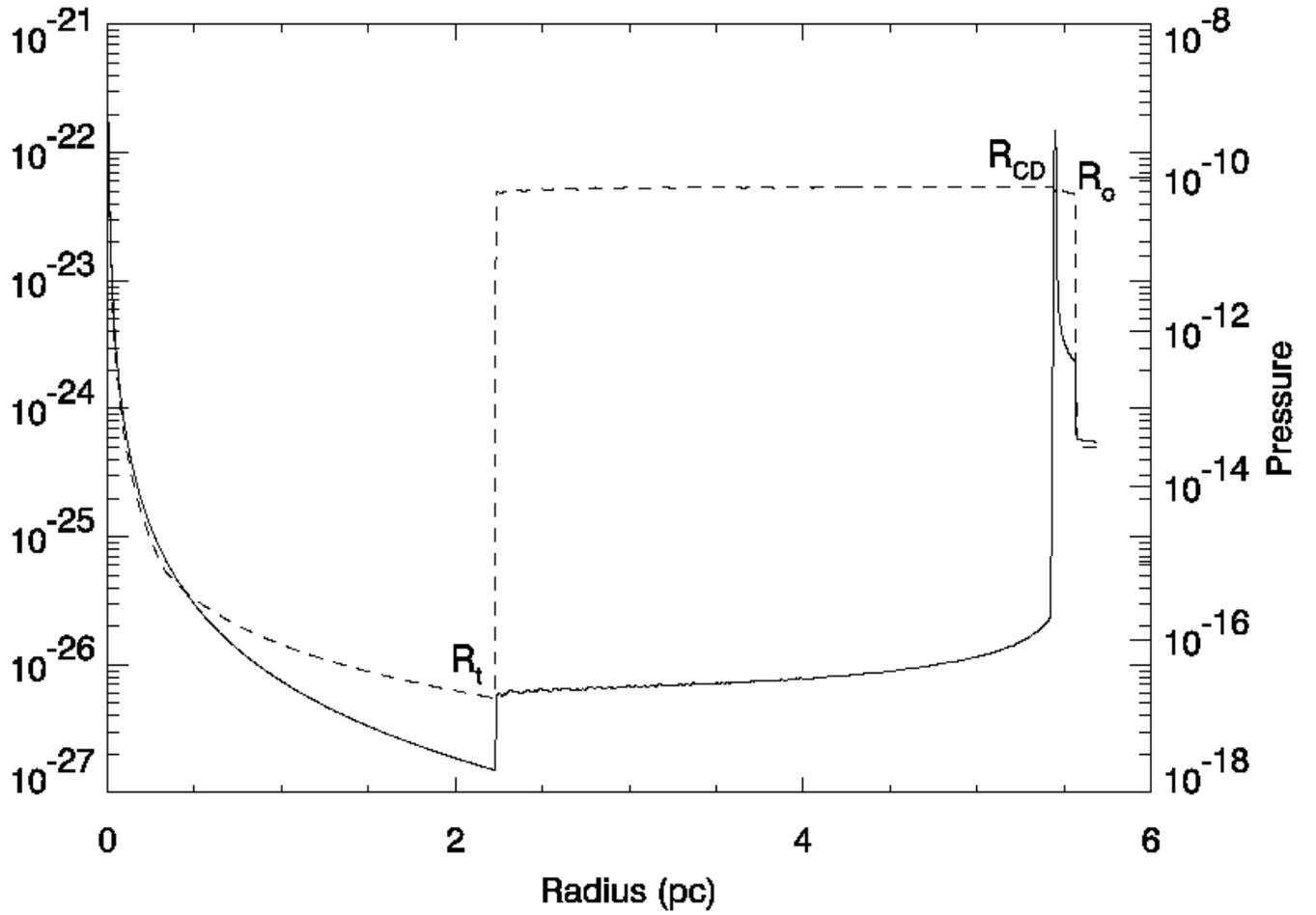}
\caption{Density (solid) and Pressure (Dashed) profiles showing the
structure of a wind-blown bubble. R$_{\rm t}$ refers to the wind-termination
shock, R$_{\rm o}$ to the outer shock, and R$_{\rm CD}$ to the radius of the
contact discontinuity. 
\label{fig:bub}}
\end{figure}

The interaction between the wind from a star and the surrounding
medium, be it the ISM or the wind from a previous epoch, can lead to
the formation of wind-blown cavities surrounded by a thin, dense, cool
shell. The structure of these wind-driven bubbles was first elucidated
by \citet[][hereafter W77]{wmc77}, and further explored by
\citet{km92a,km92b}. In the simplest case, that of mass-loss from a
fast wind with constant parameters interacting with a slower wind (or
the ISM) with constant parameters, a self-similar solution can be
obtained. Figure \ref{fig:bub} shows the pressure and density
structure within a wind-blown bubble in this case. Going from left to
right we identify four regions a) the freely-flowing fast wind b) the
shocked fast wind c) the shocked ambient medium and (d) unshocked
ambient medium. An outer or forward shock (R$_o$) separates the
shocked and unshocked ambient medium, and an inner shock,
alternatively referred to as a reverse or wind-termination shock
(R$_t$), separates the shocked and unshocked fast wind. The shocked
ambient medium is separated from the shocked fast wind by a contact
discontinuity (R$_{CD}$). The example shown is that of an
'energy-conserving' structure. The interior of this structure consists
mainly of a low-density, high pressure and therefore high-temperature
region. The isotropic pressure of this region is responsible for
driving the outer shock wave. The temperature could be high enough to
be visible in X-rays, but the emission measure is usually low, so not
many bubbles have been observed with present-day X-ray telescopes.

For a wind with constant properties expanding into a constant-density
ISM, W77 showed that a self-similar solution could be obtained for the
evolution of the contact discontinuity with time, namely R$_{CD}
\propto (L_w/{\rho})^{1/5}\; t^{3/5}$. Here $L_w$ is the mechanical
wind-luminosity $L_w = 0.5 \dot{M} {v_w}^2$. If the fast wind expands
into a slower wind from a previous epoch, the result becomes R$_{CD}
\propto (L_w/{\rho})^{1/3}\; t$. Thus the shell expands with constant
velocity. 

The above description has ignored many other contributing factors.
The wind properties are unlikely to be constant throughout the
evolution, thus giving rise to multiple shells. UV photons from the
hot stars will ionize the medium around the star, thus altering the
dynamics. Factors such as conduction and/or magnetic fields may change
the interior dynamics. Heat conduction in particular has been cited as
one of the reasons for lowering the temperature in the interior of the
bubble. Multi-dimensional effects such as hydrodynamic instabilities
can alter the evolution \citep{db98}, as we shall see in a future
paper. The presence of instabilities can result in considerable
turbulence within the interior of the bubble, breaking the
approximations of isotropic pressure and a radial velocity field.

The results above refer to a homogeneous medium. In general the ISM is
known to be inhomogeneous and clumpy. However \citep{mvl84} have shown
that an early type main-sequence star may homogenize the medium up to
a radius

$$ R_h (t_{ms}) = 56 {n_m}^{-0.3}~ {\rm pc} $$ 

\noindent
where $ t_{ms} = 4.4 \times 10^6~S_{49}^{-0.25} $ yr is the main
sequence lifetime of stars of spectral type B0 to O4, S$_{49}$ is the
ionizing flux in units of 10$^{49}$ photons/sec, and $n_m$ is the mean
density the cloud gas would have if it were homogenized. Therefore we
expect that the medium around an early type star can be treated as a
smooth medium at least for distances up to about 50 pc from the
star. In later papers we will consider the effects of a clumpy medium.

Thus far we have considered only spherical circumstellar bubbles. The
circumstellar structures around many massive stars, especially WR
stars and/or Ofpe/W9 stars, do fall in this category. However many
nebulae around Luminous Blue Variable (LBV) stars show distinct
evidence of bipolarity \citep{w01}. The bipolar structure around SN
1987A has been clearly mapped using light echoes \citep{ckh95}. It is
unclear what additional ingredients contribute to the formation of
these aspherical nebulae. A common explanation tends to invoke an
asymmetry in the surrounding medium, such as an equatorial disk, which
inhibits the expansion of these nebulae along the equator but allows
for free expansion in the polar regions, leading to bipolarity. An
alternative explanation attributes the asphericity not to the external
medium but to the wind from the star itself. For example a rotating
star may give rise to a wind that is faster in the polar regions and
slower in the equatorial regions \citep[see for example][]{do02}. In
this paper we consider the interaction of SN shock waves with
spherical nebulae. In future papers we will consider how the
interaction changes due to deviations from sphericity.

\begin{figure}
\includegraphics[angle=90,scale=0.85]{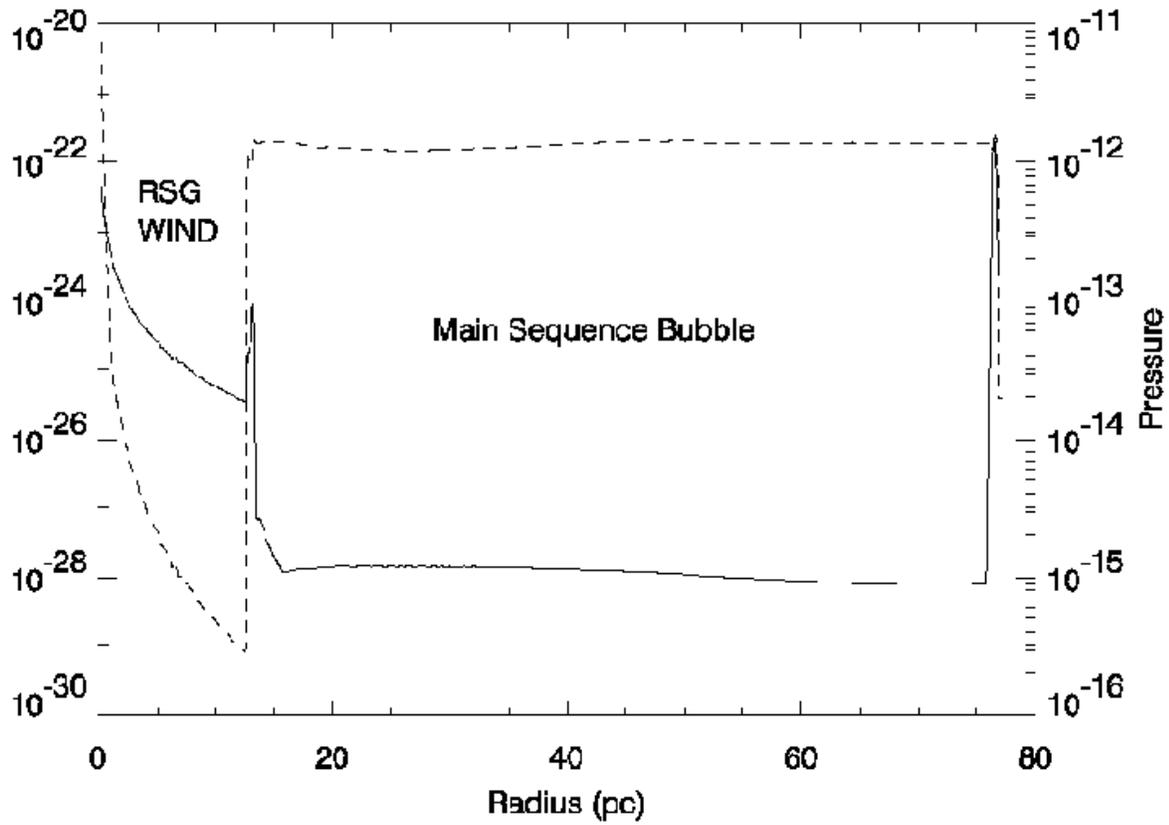}
\caption{Density (solid) and Pressure (Dashed) profiles for the medium
around a RSG star. The main-sequence wind formed a wind-blown bubble
around the star, whereas the subsequent very high density RSG wind
created a new pressure equilibrium.  \label{fig:rsgbub}}
\end{figure}

It is clear that the circumstellar bubble results from the impact of a
fast wind on a slower ambient medium. However in many cases the
situation may be reversed, i.e. it is a slower wind that impacts on a
faster wind from a previous epoch. This may very well be the general
case for RSG winds, whose velocity is about 20 km/s, and which may
follow a much faster main-sequence wind. Such a situation is
illustrated in Fig \ref{fig:rsgbub}, which illustrates a snapshot from
a simulation in which a MS wind is followed by a RSG wind. The figure
shows the density and pressure profiles close to the end of the RSG
stage, as the RSG wind creates a new pressure equilibrium. Unlike the
case illustrated above, no wind-driven bubble is created, but the wind
is almost freely expanding all the way to the interaction region with
the main-sequence wind, from which it is separated by a contact
discontinuity. The evolution of the SN ejecta in this medium would be
described by the self-similar solutions for power-law ejecta evolving
in a power-law ambient medium (see below), up until the time the
ejecta collided with the shell. Given the low wind velocity
(v$_{RSG}$) and a maximum RSG lifetime ($t_{RSG}$) of about 100000
years, the interaction of SN ejecta with a RSG wind can last for about

\begin{equation}
{\rm t}_{int} = 400~ \left({t_{RSG}}\over{100,000 ~{\rm
yrs}}\right)\left({v_{RSG}}\over{20~{\rm
km/s}}\right)\left({v_{SN}}\over{5000~{\rm km/s}}\right)^{-1}~{\rm years}
\end{equation}

\noindent
where we have used an average velocity of 5000 km/s to take
deceleration of the ejecta by the dense RSG wind into account. For
higher mass-stars the RSG phase may be more short-lived, and the
interaction time will decrease. After this period the remnant will
continue to evolve in a low-density wind-blown bubble formed by the
main-sequence star, until it collides with the dense shell.

\section{ The SN Ejecta}
\label{sec:snejecta}

We are dealing with young supernovae, which are still in an
ejecta-dominated stage. To describe the supernova evolution from the
early stages, a prescription for the profile of the ejected material
is required. This profile is established very early on, in the first
few days after the explosion. The velocity of the ejected material
tends towards free-expansion, with $v=r/t$ as expected. The density in
free-expansion at a particular velocity decreases as t$^{-3}$
\citep{cf94}.

The density distribution of the ejected material depends on the
structure of the pre-SN star. In core-collapse SNe where the explosion
energy is all contained in the center, the distribution of ejected
material depends on whether the envelope is convective or radiative
\citep[][hereafter MM99]{mm99}, and in the former case on the
composition. The radiative case is more typical of W-R stars and other
very massive stars, whereas RSGs are better described by the
convective case. Analytical and numerical calculations
\citep[][MM99]{cs89} have shown that the ejecta density can be
reasonably well represented as a power-law in the outer parts of the
ejecta ($\rho_{SN} \propto v^{-n}$), with a flat or shallower
distribution ($n < 3$) below a certain velocity $v_t$. MM99 find that
the RSG distribution can be described by a power-law with an exponent
n=11.7, whereas the convective case is better described by a slightly
shallower power law n=10.2. We note that these are approximations,
although good ones; in general the power-law may vary over the ejecta
structure. In fact \citet{dc98} showed by inspection of the numerical
models that the ejecta in Type Ia SNe could best be described by an
exponential profile.

The interaction of power-law ejecta ($\rho_{SN} \propto v^{-n}
t^{-3}$) with a power-law surrounding medium ($\rho_{CS} \propto
r^{-s}$) can be described by a self-similar solution \citep{c82},
wherein the contact discontinuity evolves with time as R$_{CD} \propto
t^{(n-3)/(n-s)}$. Such a formulation has often been used in the past
to describe SN evolution in stellar winds.

In this work we assume that the ejecta are described by a power-law,
with a power-law exponent of 9. This is close enough to the values
suggested by MM99, while also being the value suggested for the ejecta
of SN 1987A. The mass of material ejected in the explosion is taken as
10 $\msun$. Our values are different from those used by Tenorio-Tagle
and collaborators in all their publications \citep{tbf90}. They used a
SN density profile decreasing as r$^{-3}$, and a total mass of ejected
material of 4 $\msun$. Their ejecta density profile is too shallow
compared to the work of MM99. Besides, a value of $n > 5$ ensures a
finite mass and energy for the ejecta. The difference between the
density profile used herein and that employed in \citet{tbf90} is that
their shallower profile results in a considerable amount of the
ejected mass being present at large velocities. We have used a higher
value for the ejected mass on the assumption that many massive stars,
especially in the middle-mass range of 10-30 $\msun$, would leave
behind a compact remnant and eject most of their remaining mass in the
explosion. This is an arbitrary number, and in future papers we will
compute models with a smaller value of the ejected material. As
discussed below, the remnant evolution is not so much a function of
the ejected mass itself as of the ratio of the mass of the dense
circumstellar shell to the ejected mass. The evolution to the Sedov
stage however will depend on the amount of material ejected (see \S
\ref{sec:disc}).

\section{Thin-Shell Model}
\label{sec:ts}

        In this section, we develop an analytical thin shell model for
the evolution of the supernova-circumstellar shell interaction.
Although some approximations are necessary, the model shows the
dependence on the parameters and gives general insight into the
evolution.

There are two main approximations.  First, the circumstellar shell
plus swept-up external gas can be regarded as a thin shell with
thickness $\Delta R$ much less than radius $R$.  This approximation
should be adequate for the early evolution, but will break down as the
outer shock front moves away from the shell position.  Second, the
supernova energy interior to the shell can be regarded as giving a
region of constant internal energy.  This approximation is inaccurate
during the early evolution when there are pressure waves and shock
waves moving through the interior, but should become accurate later in
the evolution once complete thermalization of the ejecta energy has
occurred.  A final approximation is that the expansion is spherically
symmetric; that is, instabilities do not have a major effect on the
evolution.

        We consider the circumstellar shell to have a mass $M_{sh}$
and to be surrounded by a cold medium with density distribution
$\rho_s=\rho_a (R/R_i)^{-s}$, where $\rho_a$ is a constant, $R_i$ is
the initial shell radius, and $s$ is a constant $<3$.  The cases of
most interest are $s=0$ (constant density interstellar medium) and
$s=2$ (circumstellar gas from steady mass loss).  The mass
conservation equation is
\begin{equation}
{dM\over dt}=4\pi R^2\rho_s V,
\end{equation}
where $M$ is the shell mass including the swept up mass and
$V$ is the shell velocity.
The mass equation can be integrated to yield
\begin{equation}
M=M_{sh}[1+m(y^{3-s}-1)],
\label{eq:ms}
\end{equation}
where $m=M_i/M_{sh}$ is a constant, $M_i$ is the mass that the
surrounding medium would have if it extended from $r=0$ to
$R_i$, and $y=R/R_i$.

The shell initially receives an impulse when it is hit by the
supernova shock front.  CL89 have described some of the details of
this interaction, allowing for a steep outer density profile in the
supernova ejecta.  Here, we make the approximation that the initial
shell velocity from this interaction is $V_i=(\alpha
p_{sn}/\rho_{sh})^{1/2}$, where $\alpha$ is a constant, $p_{sn}=E/2\pi
R_i^3$, $E$ is the supernova energy, and $\rho_{sh}$ is the initial
density in the shell.  The initial pressure in the shell, $p_i$, for
the shell evolution is then given by $E=2\pi R_i^3 p_i+M_{sh}V_i^2/2$.
We find it convenient to use a dimensionless velocity variable
\begin{equation}
\label{eq:dimvel}
w^2={V^2\over mp_i/\rho_a}.
\end{equation}
With this definition, the initial shell velocity is
\begin{equation}
\label{eq:velinit}
w_i=\left[{{m \rho_{sh}}\over \alpha \rho_a}-{1\over (3-s)}\right]^{-1/2}.
\end{equation}

The equation of motion of the shell is
\begin{equation}
M{dV\over dt}=4\pi R^2(p_{int}-\rho_s V^2),
\end{equation}
where $p_{int}$ is the interior pressure and evolves as $p_i y^{-5}$
due to adiabatic expansion.  From the above discussion and
substitutions, the equation can be written as
\begin{equation}
{d w^2\over dy}={2(3-s)y^2 (y^{-5}-my^{-s} w^2)\over [1+m(y^{3-s}-1)]}.
\end{equation}
This equation can be solved using standard techniques to give:
\begin{equation}
\label{eq:wfinal}
w^2={w_i^2+(3-s)(1-m)(1-y^{-2})+2m(3-s)(1-s)^{-1}(y^{1-s}-1)\over
[1+m(y^{3-s}-1)]^2},
\end{equation}
where $s\not= 1$. If $s$ = 1, such as may occur in a region consisting
of molecular clouds \citep{cbd85, s90}, then we get 
\be
 {w^2} = {{w_i}^{2} + 2(m-1)(y^{-2} -1) + 4m\,\ln y \over {[1 + m(y^2
-1)]^2}}\;.  
\ee

For completeness, we also show the solution for planar ($q=1$),
cylindrical ($q=2$), and spherical ($q=3$) geometry when $s=0$:
\begin{equation}
\label{eq:wall}
w^2={w_i^2+3+6my^{q/3}-3(1-m)y^{-2q/3}-9m\over [1+m(y^q-1)]^2}.
\end{equation}
The parameters here have definitions equivalent to those for the
spherical case.

\begin{figure}
\plotone{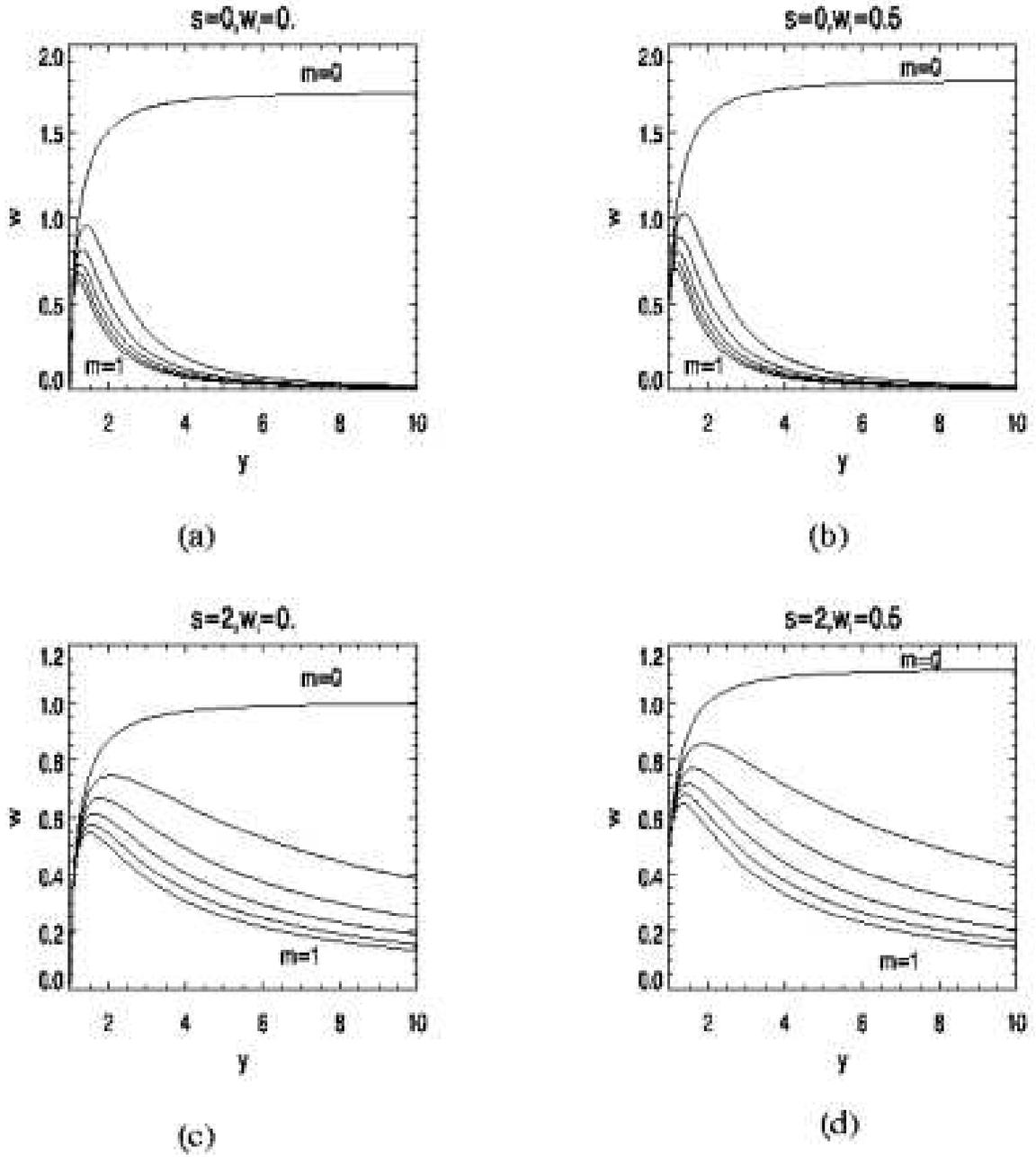}
\caption{The solution for spherical expansion with s=0 and s=2, for
two values of $w_i$, 0.0 and 0.5. The value of m varies from m=0.~to
m=1.~in steps of 0.2.\label{fig:thinsh} }
\end{figure}

The solutions for spherical expansion with $s=0$ and $s=2$ are
illustrated in Fig \ref{fig:thinsh}.  The range of $m$ is from 0 (no
surrounding medium) to 1 (mass in shell equal to mass external medium
would have if it extended from $R_i$ to $r=0$).  The $m=1$ case is
thus what would be obtained if the shell is just the swept-up external
medium, as could be the result of a fast wind from the central star
sweeping up a slower wind from a previous epoch. The $m=0$ case would
occur, for instance, if the shell were composed of mass ejected from
the central star in a low density ambient medium. Low density in this
case implies that the mass of any ejected shell is large enough, and
the ambient density small enough, that the shocked shell would not
sweep up a mass comparable to its own over many doubling times of the
radius. Mass ejections on the order of a solar mass have been presumed
to occur in Luminous Blue Variables such as $\eta$ Carinae.  If $m=0$
the shell mass remains constant with time.  Strictly speaking the
equations~(\ref{eq:dimvel}) and~(\ref{eq:velinit}) are not valid for
$m=0$, although the final solution in equation~(\ref{eq:wfinal}) still
holds.  The $m=0$ cases show little deceleration, but rapid
deceleration does occur for higher $m$. For $s=0$, it can be clearly
seen from equation \ref{eq:wfinal} that for any finite, non-zero value
of $m$, the denominator goes as $y^6$ for large $y$. This implies that
in the presence of an ambient medium, over several doubling times the
shell is invariably going to slow down to a low velocity irrespective
of the initial mass and velocity of the shell. It should be kept in
mind however that with deceleration the thin shell approximation is
expected to break down if the flow is nonradiative.

 One way to estimate the deceleration is to examine a ``braking
index'' $\Omega\equiv \ddot R R/\dot R^2$.  For a self-similar flow
expanding as $R\propto t^k$, we have $\Omega=(k-1)/k$; the thicknesses
of shocked regions for various values of $k$ can be found in
\citep{c82}.  There is a critical value of $k$ corresponding to a
point explosion blast wave for which the swept-up shell is expected to
become very broad.  The value is $k=2/(5-s)$ or $\Omega_c=-(3-s)/2$.
Once this value of $\Omega$ is reached, the shock front can be
expected to strongly separate from the shell. The braking index can
also be calculated for the shell motion, with the result for the $m=1$
case:
\begin{equation}
\Omega(s=0)=-3\left(w_i^2-6+5y\over w_i^2-6+6y\right),\quad
\Omega(s=2)=-{w_i^2+2-3y^{-1}\over w_i^2+2-2y^{-1}}.
\end{equation}
We thus find that $\Omega=\Omega_c$ for $y=(6-w_i^2)/4$ ($s=0$) and
$y=4/(2+w_i^2)$ ($s=2$); the thin shell phase is quite short-lived for
$m=1$.  At the other extreme of $m=0$, there is no deceleration, which
is expected because the ambient medium does not exert any significant
pressure on the shell, and the shell tends toward a constant velocity.
From equation~(\ref{eq:wall}), for a spherical shell this occurs when
$y^2 \gg 3$.

\section{Numerical Simulations}
\label{sec:numinfo}

The numerical simulations described herein were carried out using the
VH-1 code, a 1,2, and 3-dimensional numerical hydrodynamics code
developed by at the University of Virginia by John Hawley, John
Blondin et al. It is based on the Piecewise Parabolic Method described
in \citet{cw84}. The method is well suited to simulating compressible
flows in astrophysics, and the code has been used previously by this
author and others for astrophysical calculations. Cooling is employed
in the form of a cooling function. We use the one described in
\citet{sd93}. Cooling is necessary in order to simulate the evolution
of the CS bubble and the formation of the thin, dense, cooled shell,
but the amount of cooling is relatively insensitive to the details of
the cooling function used. Cooling is not generally significant in the
evolution of the remnant, except when the shell mass significantly
exceeds the ejecta mass (see \S \ref{sec:largel}). No contact
discontinuity steepener is used. Simulations are carried out on a
spherical-polar (r, $\theta$) grid. Two-dimensional simulations
consider only one quadrant assuming azimuthal and equatorial
symmetry. Inflow (describing SN ejecta) and outflow (describing
surrounding medium) boundary conditions are used at the inner and
outer boundary respectively, while reflecting boundary conditions are
used at the remaining boundaries in case of 2D. One advantage of this
code is that it employs an expanding grid, which tracks the outer
shock front and expands along with it. All grid zones are allowed to
expand, and no new grid zones are added (the code is not
adaptive). This feature is very useful in simulations where the
dimensions of the system change by many orders of magnitude over the
evolutionary timescale.

\section{Supernova-Circumstellar Bubble Interaction}
\label{sec:snbub}

The structure of the CS bubble as described in \S{\ref{sec:wbb}}
consists of a highly evacuated cavity surrounded by a thin, dense
shell. When the star explodes as a SN, the shock wave will interact
primarily with a low density medium, with density lower than that of
the surrounding ISM. Emission from the remnant arises mainly from the
shocked interaction region between the SN shock wave and the ambient
medium \citep{cf94}. The low density results in a much lower intensity
of emission, since emission at most wavelengths is a function of the
square of the density. Thus the emission from the remnant will be
considerably reduced compared to an explosion in the ISM, i.e.~the
remnant will be muffled \citep{m88}. In some cases explosion within a
circumstellar bubble made by a hot massive star may render the remnant
unobservable until the shock wave impacts the dense circumstellar
shell.

 The density and pressure profiles within the bubble (Figure
\ref{fig:bub}) show a double-shocked structure separated by a contact
discontinuity. When the SN ejecta collide with the freely expanding
wind, they drive a strong shock into the wind. A reverse shock is also
formed which decelerates the ejecta material.  Due to the presence of
many shocks and density discontinuities, the subsequent interaction is
quite complex, as demonstrated in analytical studies carried out by
CL89. In what follows we focus on numerical studies of this
interaction.

Most of the mass of the circumstellar bubble is contained in a thin,
dense, cool shell bordering the cavity, which suggests that this is
the essential parameter influencing the dynamics. This is verified by
our simulations and previous work by others \citep{tbf90}. The
interaction is found to depend on one parameter, the ratio of the mass
in the swept-up shell to the mass of the ejected material, which we
denote as $\Lambda$. If the bubble was formed by the interaction
between two winds, then the swept-up mass, and hence $\Lambda$, will
vary as the radius $r$ (and not $r^3$ as it would for a constant
density surrounding medium).

The role played by the $\Lambda$ parameter can be understood by
comparing to a somewhat analogous situation of an obstruction placed
in the path of a body of flowing water such as a river. A small pebble
is obviously not going to make much difference. A slightly bigger
stone may have an instantaneous impact, but the momentum of the
flowing water will quickly overrun it and soon forget that it ever
existed. A large boulder which have a much larger impact, and the
water, if it is not able to move the stone, will have to flow around
it. The flow will be diverted and its velocity altered. And finally if
the river is blocked by a dam, the flow will be halted, or at least
controlled by the obstruction.

In the same way, $\Lambda << 1$ will not have a substantial impact, as
the momentum of the ejected material will quickly destroy the
pre-existing shell. The impact of the surrounding shell begins to be
felt when $\Lambda \la 1$.When $\Lambda >> 1$ the shell can
significantly alter the supernova evolution, speed-up the remnant
evolution and perhaps even alter the various phases through which the
remnant will evolve.

We illustrate here two cases, Case 1 where $\Lambda \la 1$, and Case 2
where $\Lambda > 1$. In each case we start with the formation of the
bubble by the interaction of two winds. The initial wind was assumed
to have a high mass-loss rate ($\dot{M} = 2.5 \times 10^{-5}
\msun$/yr) and a low wind velocity (50 km/s), analogous to say a RSG
wind. The second mass loss phase was assumed to have a slightly lower
mass loss rate ($\dot{M} = 8\times 10^{-5} \msun$/yr), but a much
larger wind-velocity (2500 km/s), appropriate say for a WR wind. These
values are of course arbitrary and vary with each individual case, but
they do serve to illustrate the basic point. This assumption is akin
to ignoring a main-sequence bubble (although see Case 2 below).  This
will be an additional ingredient for the future, but the two-wind
interaction is simple and sufficient enough to understand the basic
characteristics of the SN-bubble interaction. The differences between
the bubble in the two cases is due to the different lifetime of each
wind phase.

The SN density profile is described by a power-law with exponent
$n=9$. Below a certain velocity, the density is flat in the inner
regions. The ejecta mass is uniformly taken to be 10 $\msun$ and the
explosion energy to be 10$^{51}$ ergs. The evolution of the SN shock
wave into the freely expanding wind is calculated with a
high-resolution computation. These two simulations are then mapped
onto the grid to form the initial conditions for the SN-CS bubble
interaction study.

\subsection{Case 1: $\Lambda ~ 0.14$}
\label{sec:snbub014}

We first consider a case where the mass of the wind-swept shell is
14\% of the ejecta mass. We assume that this bubble is formed within
the interstellar medium i.e.~although the medium surrounding the dense
shell is a slow wind, we assume that the density of this wind will not
fall below that of an ISM density of 1 particle/cc.

\begin{figure}
\plottwo{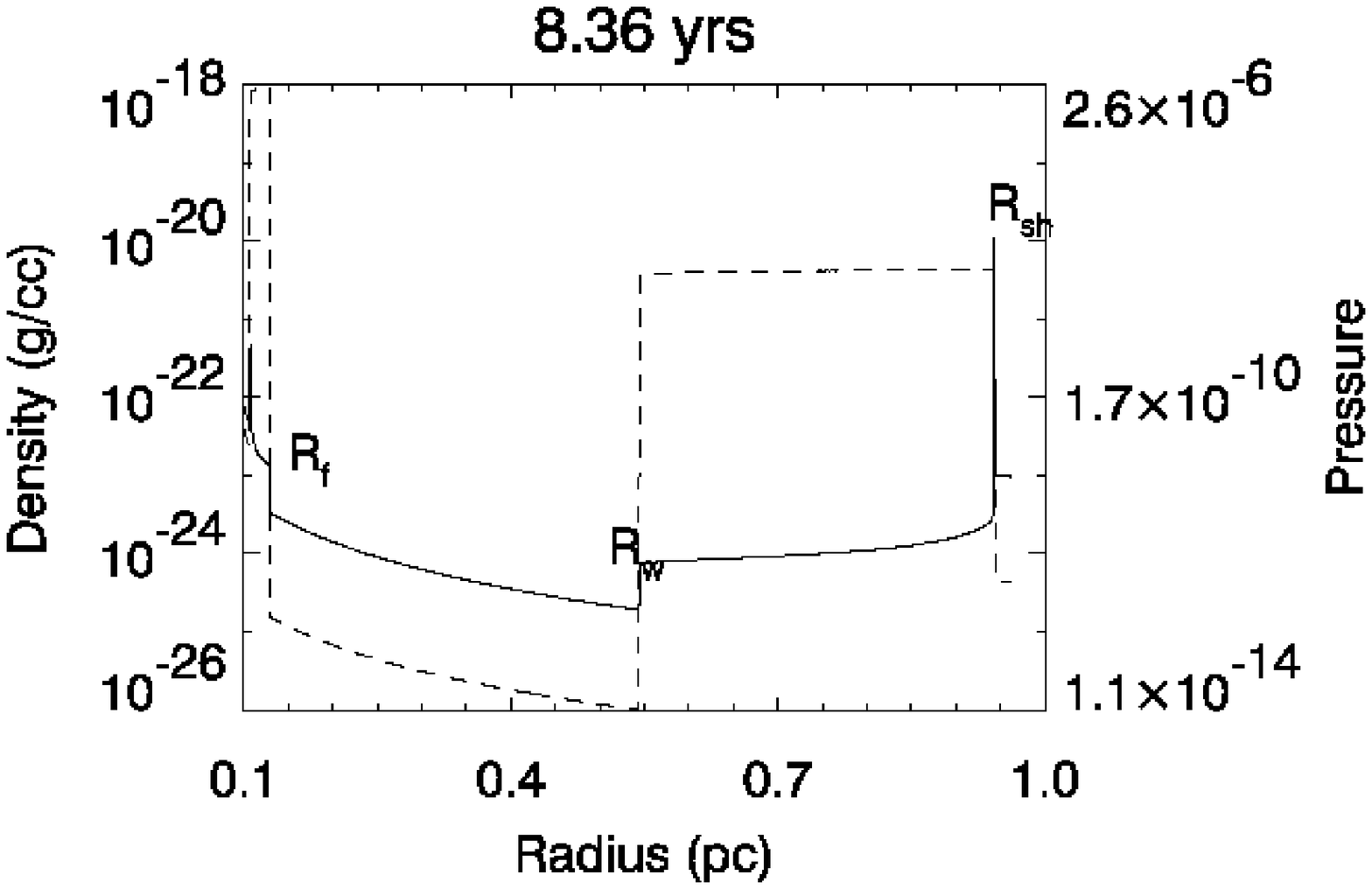}{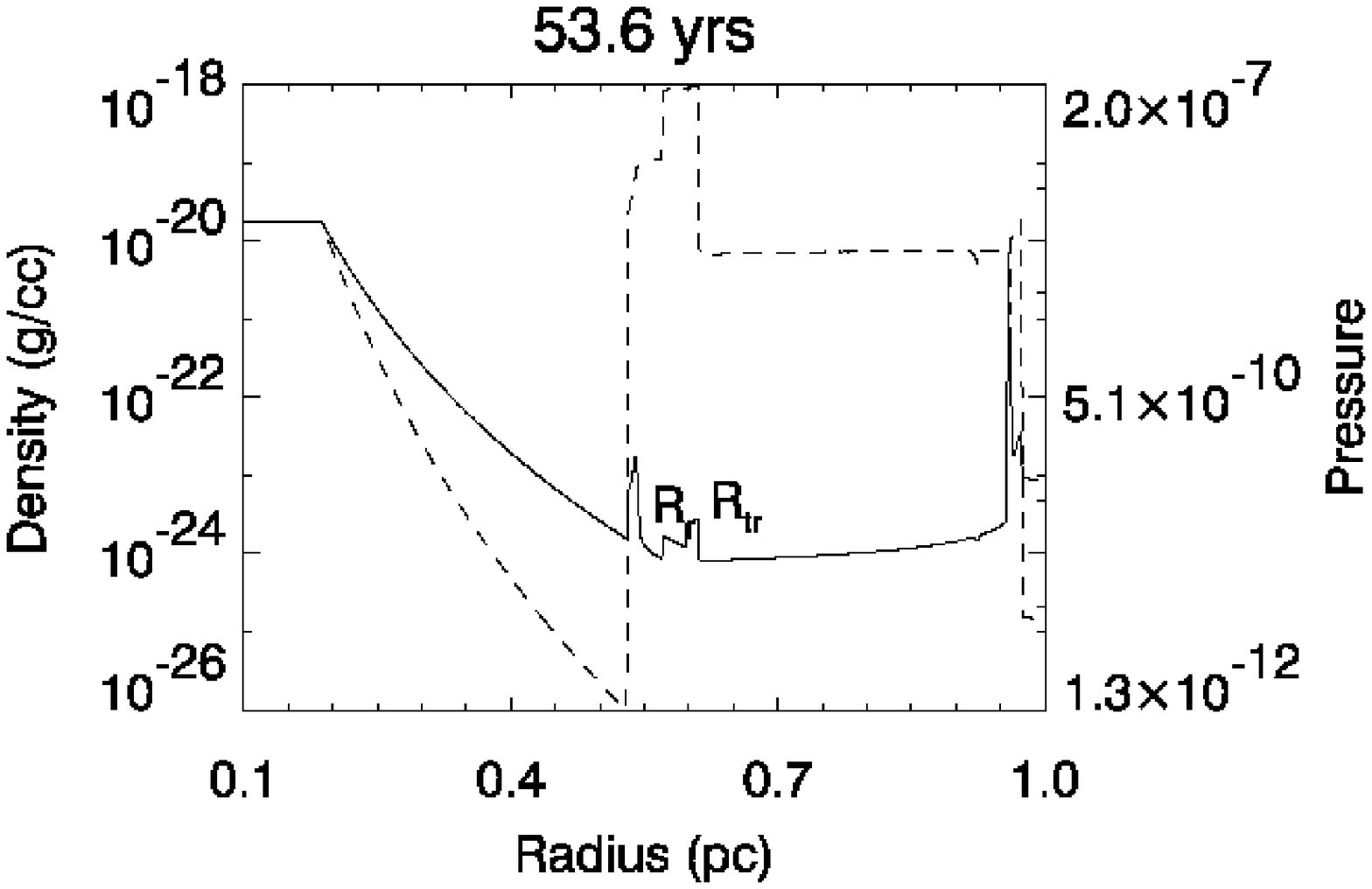}
\plottwo{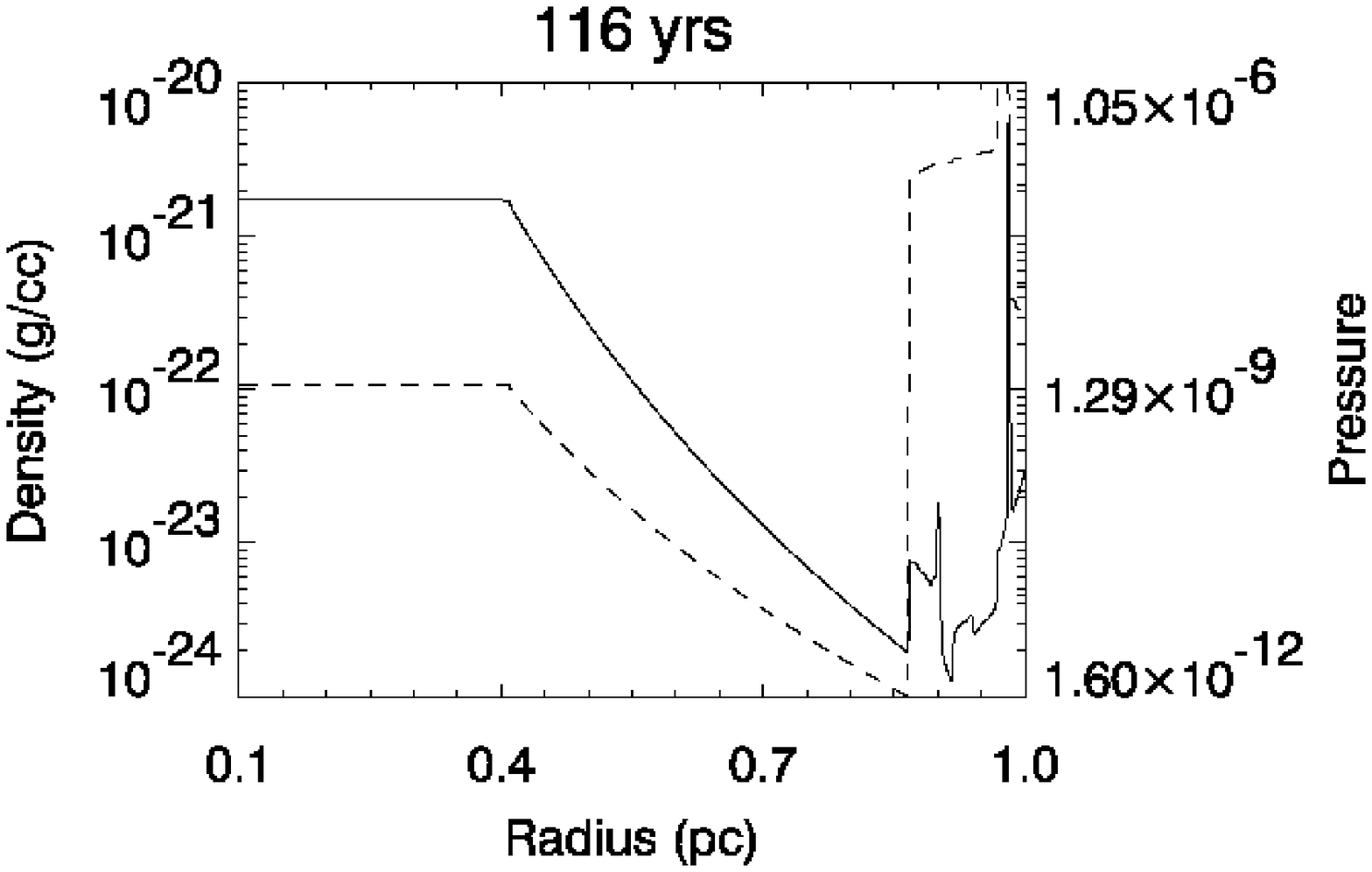}{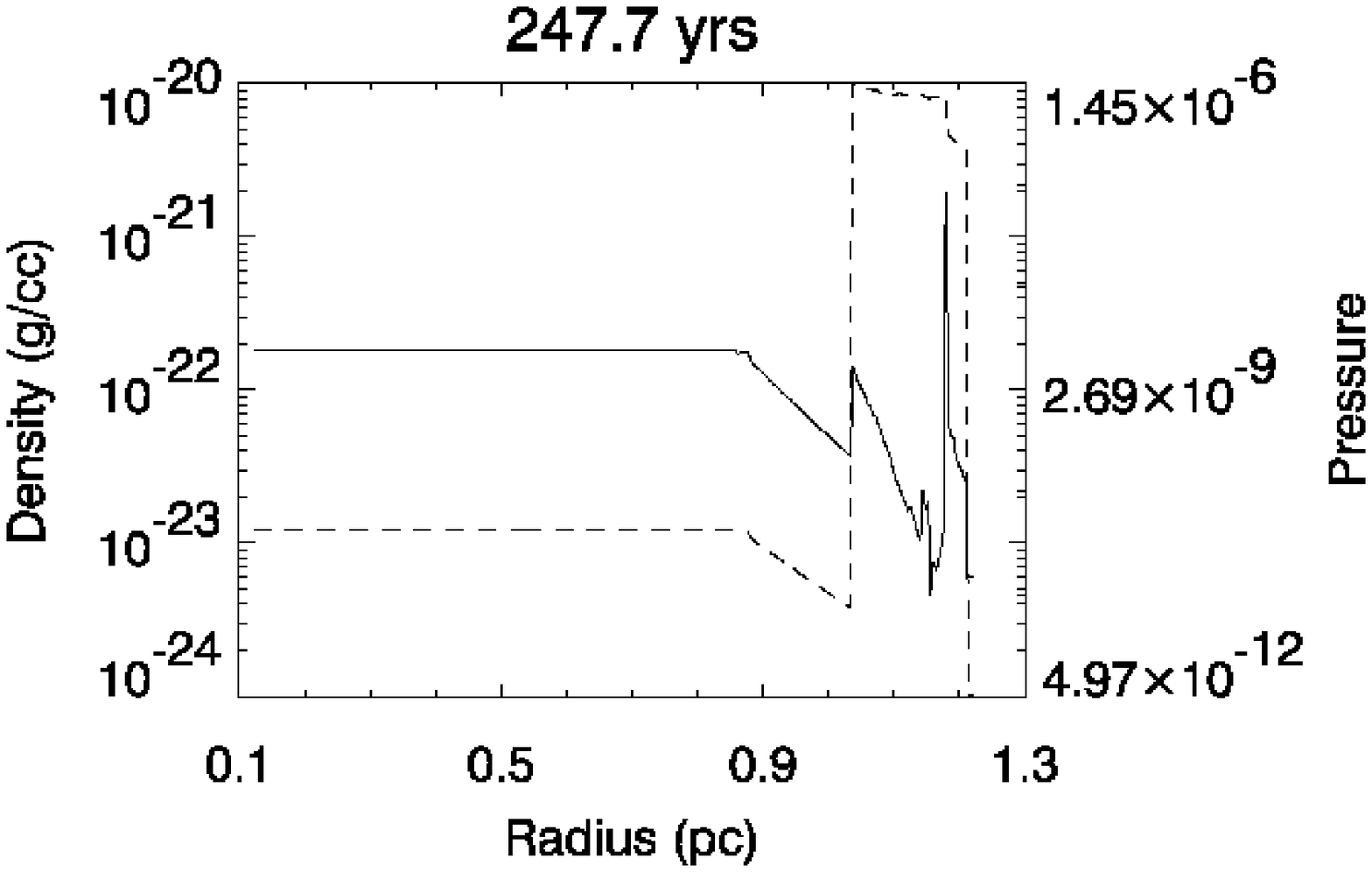}
\plottwo{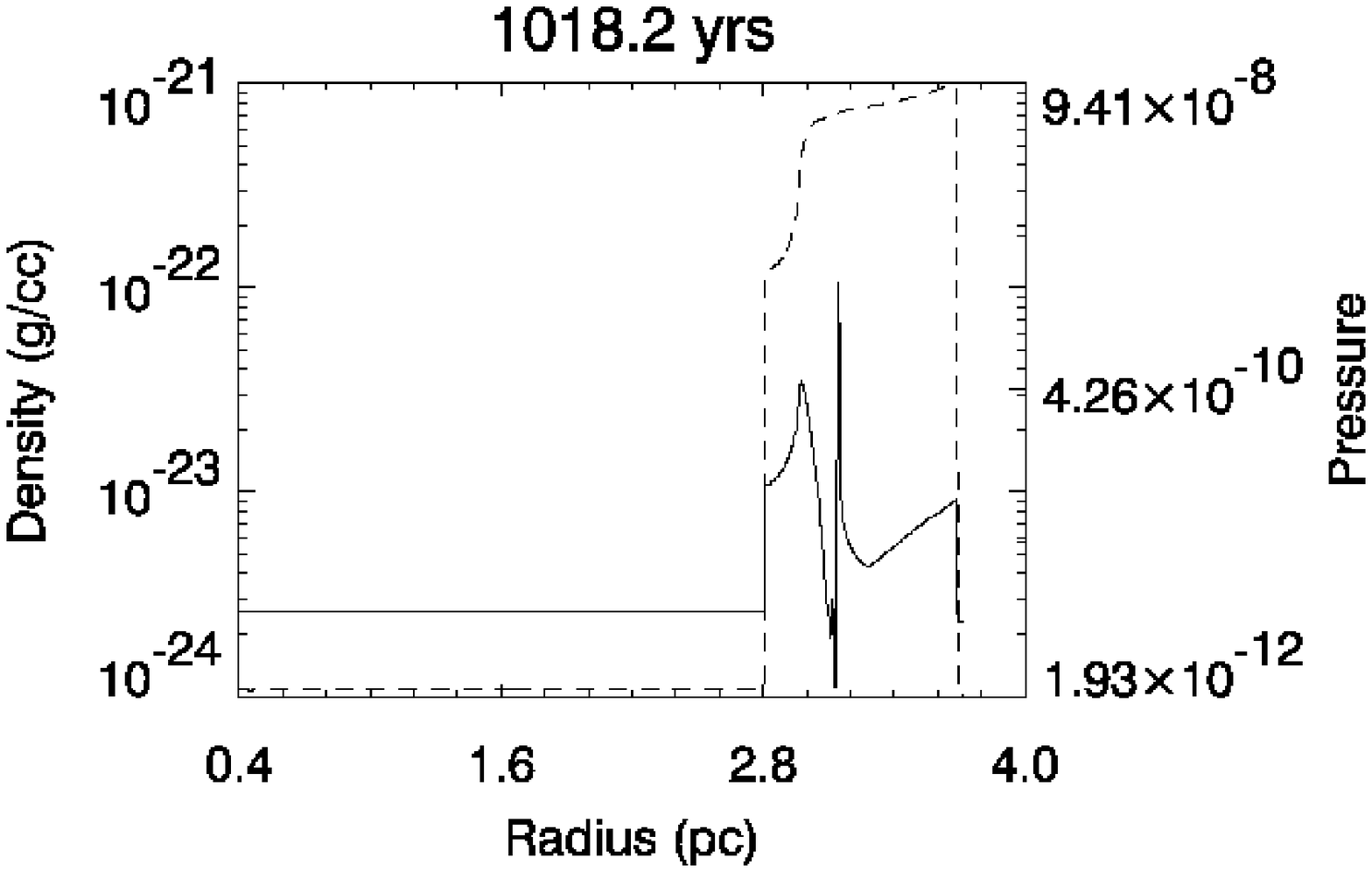}{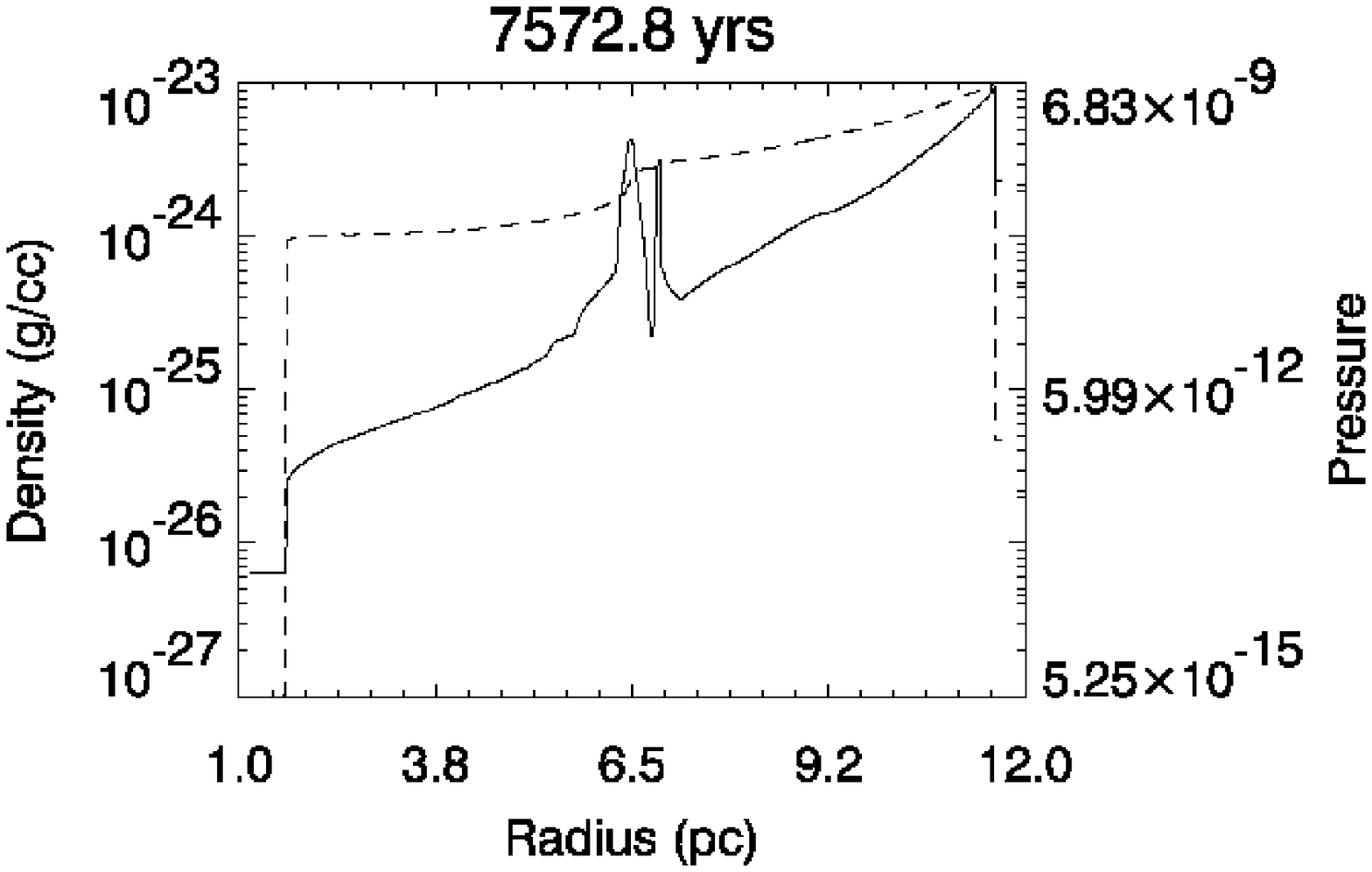}
\caption{Plots showing the density and pressure profile at various
times, for the simulation with $\Lambda = 0.14$ Solid line is density,
dashed line is pressure. The locations of the SN forward shock
(R$_f$), the wind termination shock (R$_w$) and the swept up shell
R$_{sh}$ are shown in the first frame. Refer to text for other
details. \label{fig:snbub014}}
\end{figure}

Figure \ref{fig:snbub014} shows snapshots from the evolution of the
supernova shock-wave within the bubble. The solid lines indicate
density, and the dashed lines pressure. R$_f$ indicates the SN shock
wave driven into the freely-expanding wind. R$_w$ signifies the
wind-termination shock, and R$_{sh}$ the wind-swept shell. The
simulation starts at approximately 8.36 years, which is the time that
the SN shock has been propagating in the freely expanding wind region
of the circumstellar wind blown bubble.

The expanding shock wave collides with the wind termination shock in
about 45 years, leading to the formation of a transmitted shock wave
that expands into the roughly constant density portion of the
wind-blown bubble, and a reflected shock wave that traverses back into
the SN ejecta. The collision results in an instantaneous increase in
the pressure, as is clearly visible in Fig \ref{fig:snbub014} at time
53.6 years. Note that each time the SN shock wave collides with a
shock or density discontinuity, a transmitted and reflected shock wave
pair will be formed. The increased pressure will result in an
increased temperature, leading to a rise in X-ray emission following
the impact.

The transmitted shock continues to expand in an almost constant
density medium until it reaches the wind-blown shell. Meanwhile the
reflected shock, which has a high velocity, moves back past the
original SN reverse shock and into the SN ejecta. Although the
reflected shock is moving backward in a Lagrangian sense, to an
external observer the entire system is expanding outwards in radius.

\begin{figure}
\includegraphics[angle=90, scale=0.7]{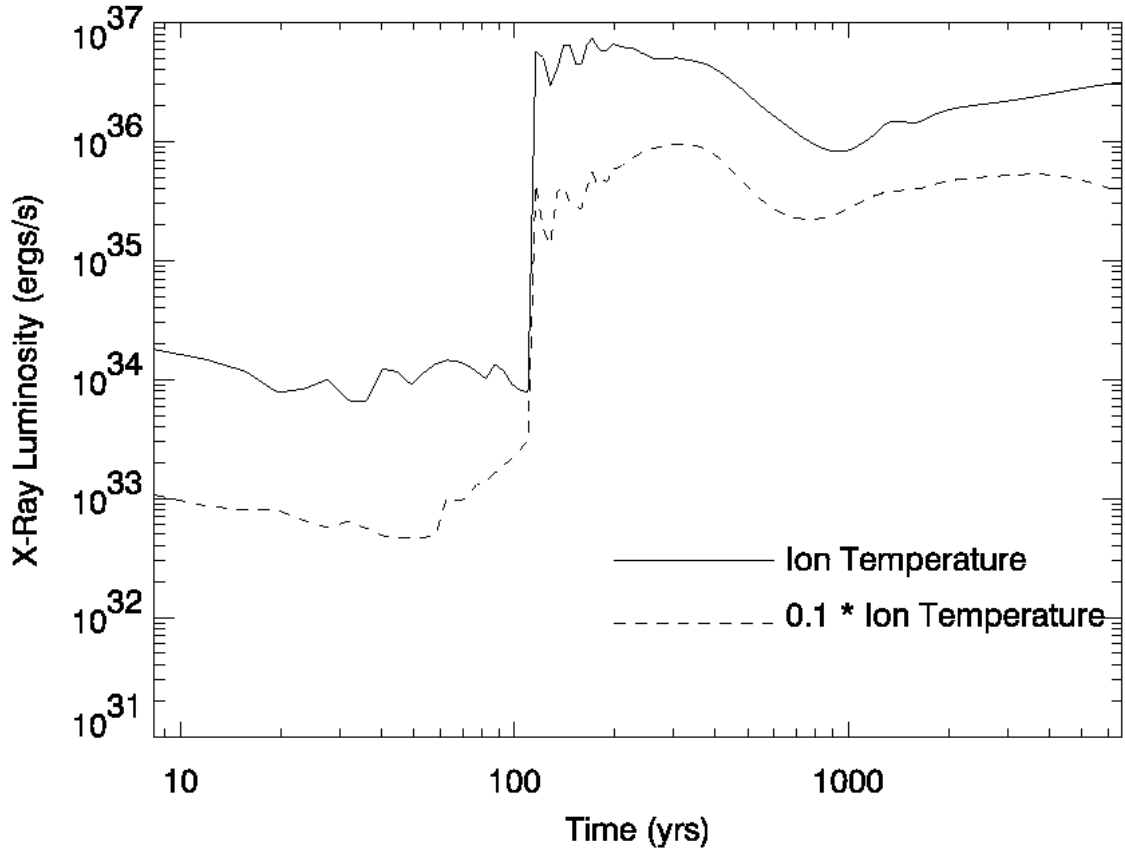}
\caption{The X-ray Luminosity During the Evolution of the remnant in
Case 1. Two curves are shown, corresponding to the temperature of the
ions within the plasma, and to a value of 10 \% of the ion temperture,
which is probably more indicative of the electron temperature in the
collisionless shock (see text).
\label{fig:xray014}}
\end{figure}

At about 110 years the forward shock (the transmitted shock from the
previous stage) collides with the surrounding dense shell. The
collision results in a compression of the dense shell, and is again
marked by a sharp rise in pressure and therefore temperature, and a
dramatic increase in the X-ray luminosity of the remnant (see Figure
{\ref{fig:xray014}}). As illustrated before, the collision results in
a transmitted shock (which is not immediately apparent as it takes
some time to emerge from the dense shell) and a reflected shock. The
high pressure generated by the interaction gives the newly formed
reflected shock a much higher velocity than the SN reverse shock. The
reflected shock therefore overtakes the reverse shock in its journey
back to the center.

The nature of the density profile is of significant interest at this
point. As can be seen in Fig \ref{fig:snbub014} at time 247 years, the
density decreases as we move outwards in radius from the reflected
shock. This will be reflected in the X-ray surface brightness profile.
However as the supernova continues to evolve, other factors come into
play: over a period of a few doubling times the remnant begins to
'forget' that the interaction ever took place. This will happen when
the mass of the swept up material becomes much larger than that of the
shell.

At about the same time, the reflected shock reaches the constant
density ejecta region. The net effect is that the SN density profile
begins to change, as can be clearly seen in Figure \ref{fig:snbub014}
at time 1018 years. The density, which was initially {\em decreasing}
from the reflected shock to the contact discontinuity, begins to {\em
increase} from the reflected shock to the contact. By a thousand years
this change is clearly visible, although it takes several thousand
years before the density profile will completely change and the SN no
longer displays any trace of the shell interaction.

It is important that this change be taken into account when computing
the emission from the remnant.  The changing density profile will
substantially alter the appearance of the remnant, as described
below. As seen in the last frame of Figure \ref{fig:snbub014}, it
takes some time for this change to become complete, when the remnant
no longer shows any traces of the interaction. In our simulation the
reflected shock converges towards the center before this occurs, but
the remnant in the last frame of figure \ref{fig:snbub014} is close to
this stage.

An interesting point to note is at time 1018 years in figure
\ref{fig:snbub014}. There is a region just behind the reflected shock,
where the {\em density is decreasing} as we move outwards in radius,
whereas the {\em pressure is increasing}. The opposing density and
pressure gradients indicate that this region will be unstable to the
Rayleigh-Taylor instability. Our multi-dimensional simulations confirm
this suggestion. The instability is of particular importance because
it occurs within the remnant, and as the shock moves inwards the
unstable region moves in and away from the forward shock. The
instability may lead to the formation of clumps and small-scale
structure in the interior of the remnant.

\begin{figure}
\plotone{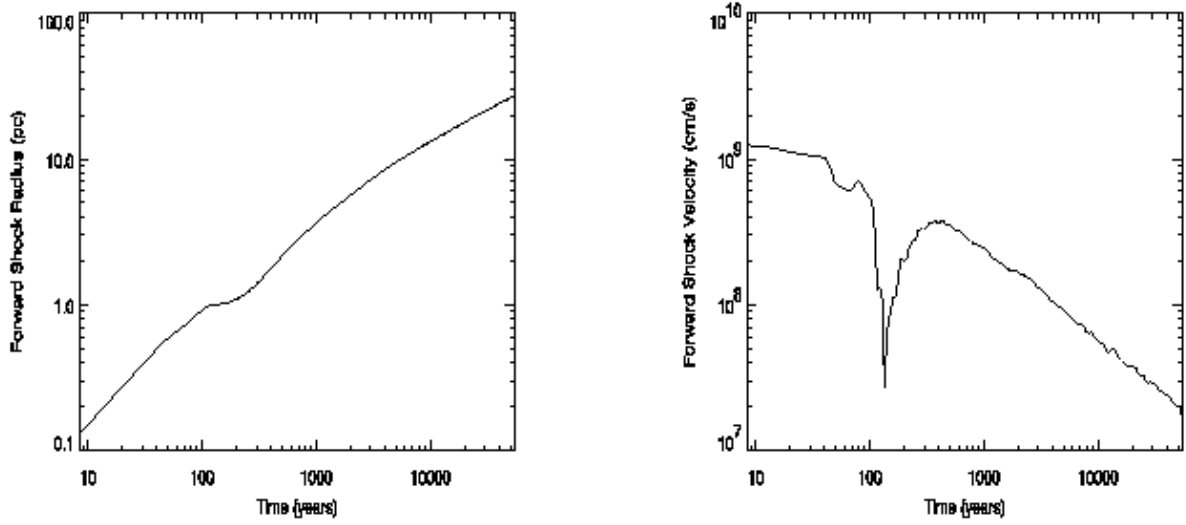}
\caption{Radius(left) and Velocity (right) during the evolution of the
SNR for Case 1. \label{fig:radvel014}}
\end{figure}

\begin{figure}
\includegraphics[angle=90, scale=0.6]{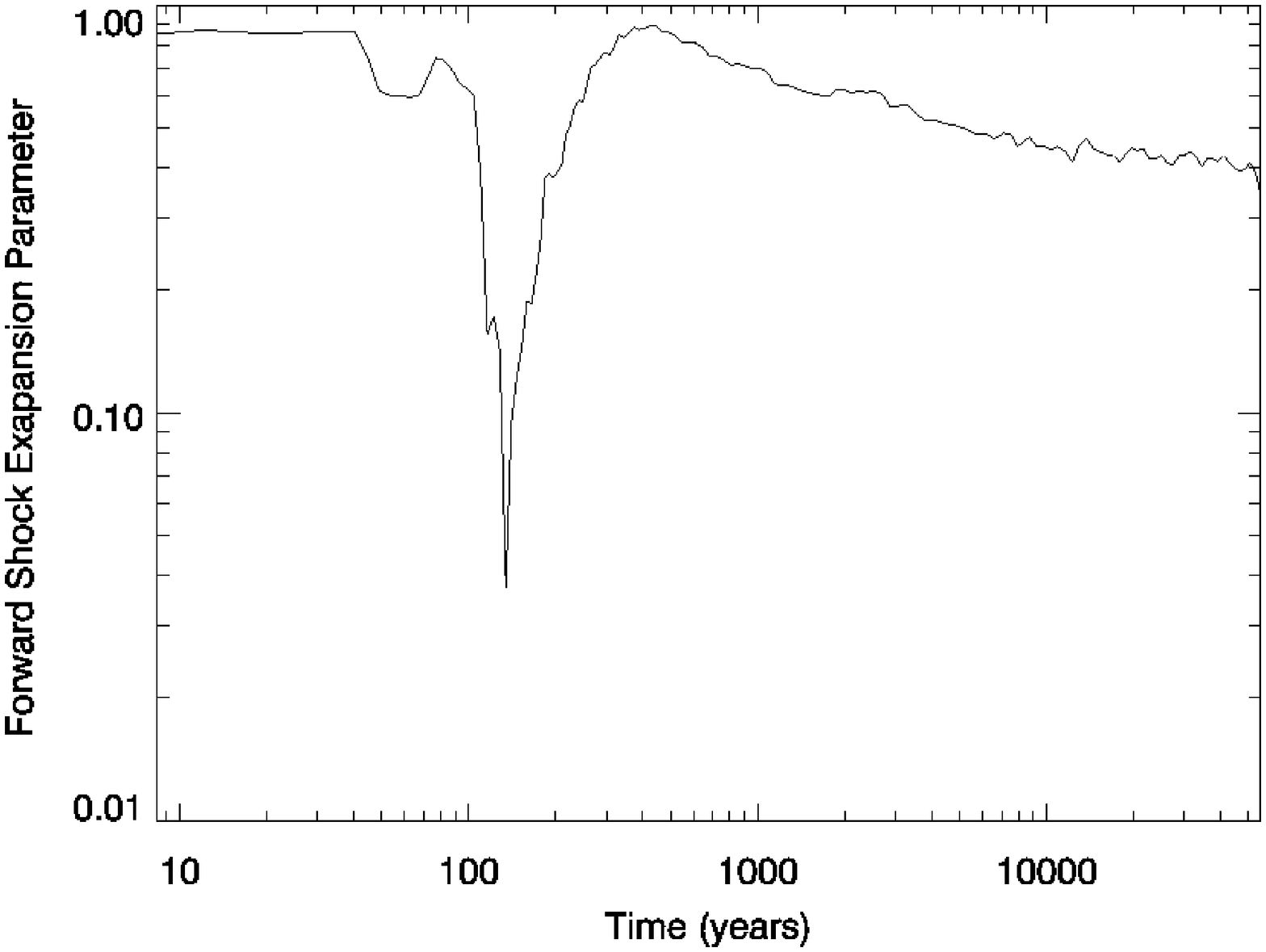}
\caption{The Evolution of the Expansion parameter for Case 1. 
\label{fig:exppar014}}
\end{figure}

The evolution of the radius and velocity of the forward shock are
shown in Fig \ref{fig:radvel014}. The expansion parameter $\delta$
(where $R \propto t^{\delta}$) is shown in Figure
\ref{fig:exppar014}. For a SN shock wave with power-law density ejecta
evolving in a medium with a power-law density structure, the value of
$\delta$ is a constant \citep{c82}. A constant value of 0.86 is
evident in the early part of this simulation, while the shock wave is
evolving in the freely expanding wind region. However this behavior
changes once the shock wave impacts the wind termination shock, as is
readily apparent in Fig \ref{fig:exppar014}. A much larger change
occurs when the shock impacts the dense wind-blown shell. The velocity
of the shock decreases considerably, its radius remains almost
constant, and hence the expansion parameter drops precipitously. After
the transmitted shock emerges from the shell, it is continually
sweeping up interstellar material, and its velocity decreases much
faster than within the bubble. As shown in Fig \ref{fig:exppar014} the
expansion parameter also gradually decreases. Once the supernova has
completely ``forgotten'' about the interaction and the reverse shock
has reached the center, we would expect the supernova to evolve to the
Sedov solution. In this case the remnant is just about approaching
this value at the end of the simulation.

 The current example, and others studied in this paper, correspond to
the m=1 case in \S \ref{sec:ts}, wherein the wind-driven shell is
fully composed of swept up material. It is difficult to determine the
precise time when the forward shock emerges from the shell, and
therefore the precise value of $w_i$, but it is low. As mentioned in
\S \ref{sec:ts}, for the m=1, s=2 case, the shock will separate from
the shell by the time the radius of the thin shell has doubled, if not
earlier, depending on the initial value of the velocity imparted to
the shell. This is clearly visible in Figure \ref{fig:snbub014}. The
initial radius of the wind-blown shell is just less than a parsec. By
the time the remnant is a 1000 years old, the shell radius has more
than tripled, and the forward shock can be seen to have already
detached from the shell. The velocity profile of the forward shock
resembles that for the m=1, s=2 case (Figure \ref{fig:thinsh}), with
an initial rise followed by a gradual decline, although the
calculation refers to the shell, not the shock wave itself.

A key observation from Fig \ref{fig:exppar014} is that the expansion
parameter varies almost continuously with time once the SN shock
impacts the wind-termination shock. This behavior can be contrasted
for evolution of the supernova in a wind or constant density medium,
where the expansion parameter is a constant for power-law ejecta. The
expansion parameter is a quantity that can be measured independent of
the distance to an object. It is used to compute the age of the
remnant, to discriminate between various models of SNR expansion, and
to determine the current phase of expansion. The varying behavior of
the expansion parameter within a bubble may cause a problem for such
determinations, because what one gets is only an instantaneous value
that is not representative of the stage of expansion.

In order to illustrate the changing morphology and appearance of the
remnant, as well as the increase in emission due to the shock-shell
interaction, we have computed the X-ray emission from the
remnant. Figure \ref{fig:xray014} shows the X-ray luminosity of the
system over time. The CHIANTI database was used for this purpose. The
X-ray luminosity computed here consists of the free-free and
free-bound luminosity, calculated over the waveband 0.2-12.4 Kev. Line
emission in X-rays is neglected. This is a good approximation for
temperatures above about 3 $\times$ 10$^7$ K, but lines may be
dominant at lower temperatures. Over most of the evolution, the
temperature in the shock interaction regions is larger than 10$^7$ K,
and the luminosity calculated here would suffice. Towards the end of
the evolution however the temperature, especially that behind the
reverse shock, falls below this value, and line emission may be
important, in which case our calculation would seriously underestimate
the total luminosity.

Almost all the X-ray emission arises from the interaction region
between the inner and outer SN shock waves. A major approximation that
is made is to assume that the electron temperature is equal to the
post-shock temperature computed by the ideal gas law
(T=$(\mu\;m_H/k)~P/\rho$). A mean value of $\mu$ is used, i.e. no
distinction is made between the mean molecular weight of the ejecta
and ambient medium. Since these are collisionless shocks, the
equilibration time due to Coulomb collisions can be large (a few
hundred to thousand years) and therefore the electron and ion
populations are not expected to be in equilibrium. Most of the
post-shock energy goes into heating the ions. Electrons are
subsequently heated by plasma processes that raise the temperature,
until temperature equilibration between ions and electrons
occurs. Thus it is possible that, especially in the early stages of
evolution, the electron temperature is much lower than the ion
temperature. In Figure \ref{fig:xray014} we therefore show two curves
for the X-ray luminosity, one using the kinetic temperature (solid
line), and one which assumes that the electron temperature is only
10\% of the kinetic temperature (dashed line). It is likely that in
the early stages the luminosity will start out closer to the lower
curve, while as the evolution proceeds the upper curve will be more
representative of the actual luminosity.

The qualitative behavior of the two curves is similar, and the main
features are seen in both. In general we note three regions, an
initial low-luminosity X-ray region, a sharp increase in the X-ray
luminosity when the shock-shell impact takes place, and then a roughly
constant high luminosity region following the shock shell impact. The
impact results in an increase of almost two to three orders of
magnitude in the X-ray luminosity. Although not computed here, a
similar rise in luminosity is expected at radio wavelengths. The steep
rise in emission and higher luminosity subsequent to the emission are
characteristic of shock-shell impact.

In order to illustrate the changing appearance of the remnant, we have
also computed approximate X-ray surface brightness profiles to
demonstrate the X-ray appearance of the remnant at various
times. Since the emission depends on the square of the density and
square-root of the temperature, we have computed the integral of this
product over the line of sight. This gives us a fair idea of what the
X-ray surface brightness profile would look like, up to a normalization
factor. All the profiles are normalized to an arbitrary factor, with
the same factor being used in both cases, so that the relative
intensities can be easily compared.  Only emission processes have been
included, no attempt has been made to take any absorption into
account. Since the intra-cavity density is so low, any absorption of
the emission that occurs will happen in the dense shell only.

\begin{figure}
\includegraphics[scale=0.87]{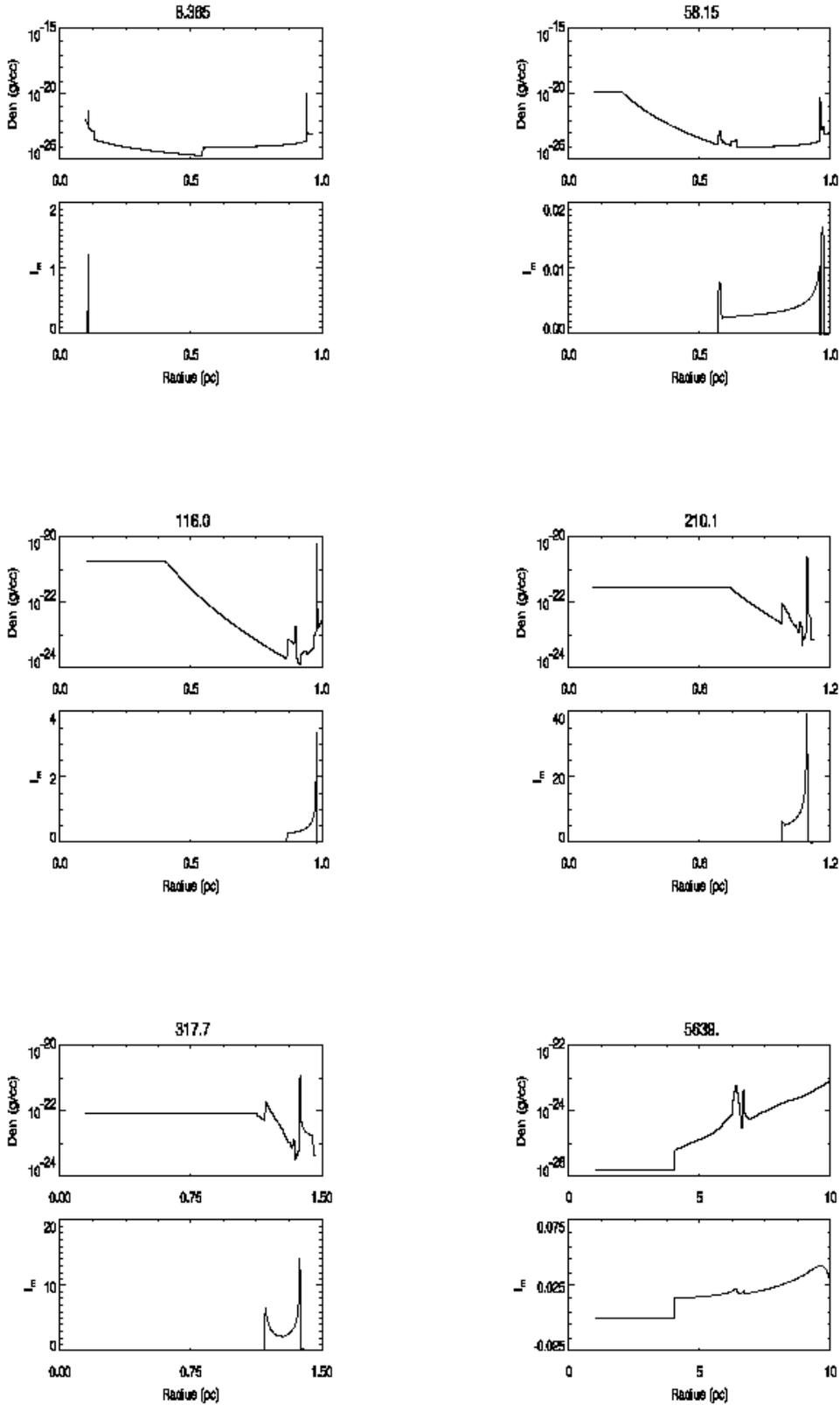}
\caption{Plots showing the evolution of the density (upper panel) and
X-ray surface brightness (lower panel) at various times, for the
simulation with $\Lambda = 0.14$
\label{fig:xraysb014}}
\end{figure}

Figure \ref{fig:xraysb014} shows the density profile, and
corresponding surface brightness profile, at various times during the
evolution. Initially only the SN shock wave is visible in X-rays, as
expected. By the time the SN shock reaches the wind-termination shock
of the bubble, the X-ray emission from the post-shock region in the
low-density ejecta is reduced so much that the wind-blown bubble
becomes the dominant X-ray source. The interaction of the SN shock
with the wind-termination shock leads to a small spike in the X-ray
emission. A much larger jump in X-ray intensity is seen when the SN
shock collides with the dense shell, and the transmitted shock
emerges. The remnant then appears limb-brightened in X-rays. As the
reflected shock traverses the power-law ejecta region, the X-ray
emission from the reflected shock also begins to play a major role,
and the SN shows a double-shelled appearance in X-rays. However soon
after the profile changes due to two factors - the remnant begins to
``forget'' the interaction, and the reflected shock reaches the inner
constant density core of the ejecta. The remnant no longer shows a
double-shelled appearance, but shows a decreasing X-ray intensity as
we move towards the center of the remnant. Note however that the
timescale between frames 5 and 6 is large, a factor of 15 doubling
times, which indicates that for a considerably large time the young
supernova remnant will show a double-shelled appearance in
X-rays. Certainly remnants with a double-shelled appearance in X-rays
are not uncommon, one example being Cas A.

\subsection{Case 2: $\Lambda$ = 3.7}

In the second instance we consider a situation where the mass in the
swept-up shell is 3.7 times that of the ejecta. The ratio $\Lambda$ is
about 25 times larger than in Case 1.  In order to produce a large
mass shell we assumed that the wind-wind interaction continues for a
very large time, while the surrounding wind density continues to
decrease. This implies that the interaction occurs in a lower density
medium outside the dense shell, perhaps a pre-existing bubble from a
previous stage. The density of the slower wind was allowed to fall to
a very low value of about 3 $\times$ 10$^{-3}$ in the simulation,
before a floor to the density was set. Note that although this may
appear very low, the surroundings of massive stars are complex and
very low density regions are not uncommon \citep{m88}. If there was a
surrounding main-sequence bubble from a massive star its density could
easily be less than 10$^{-3}$ particles/cc.

As the mass of the shell increases relative to that of the ejecta, the
energy transmitted by the ejecta to the shell also increases. Thus the
SN shock wave can lose a substantial amount of energy upon interaction
with the dense shell, and the transmitted shock is correspondingly
weakened. The shock wave may also take a larger time to emerge from
the shell.

The large shell mass results in a significant increase in pressure as
the SN shock impacts the shell. This high pressure behind the
reflected shock can result in significantly large velocities for the
reflected shock wave, which subsequently expands in a low density
medium. The reflected shock can attain very high velocities, and
reaches the center in a short time compared to the emergence of the
transmitted shock.

\begin{figure}
\plottwo{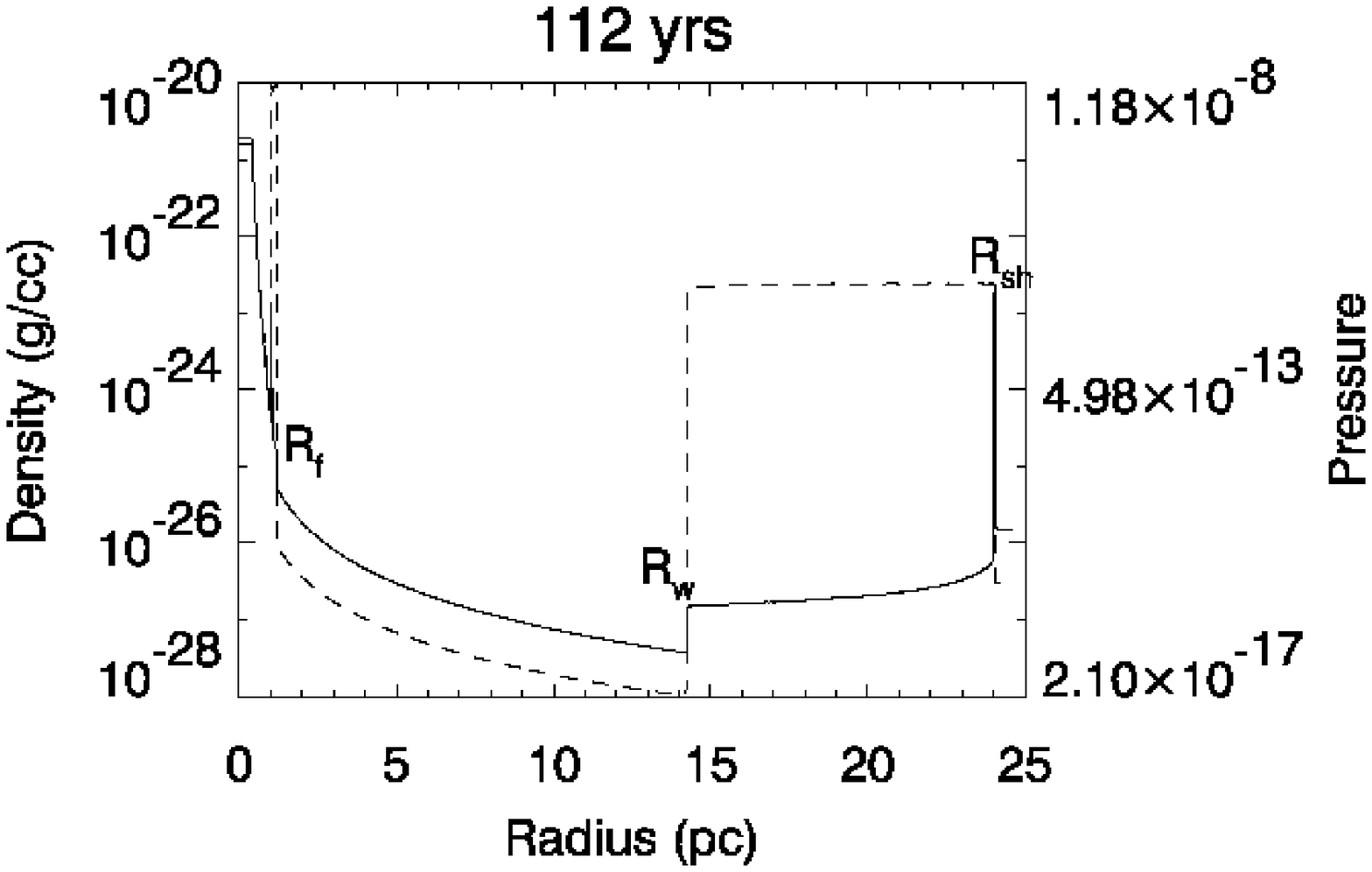}{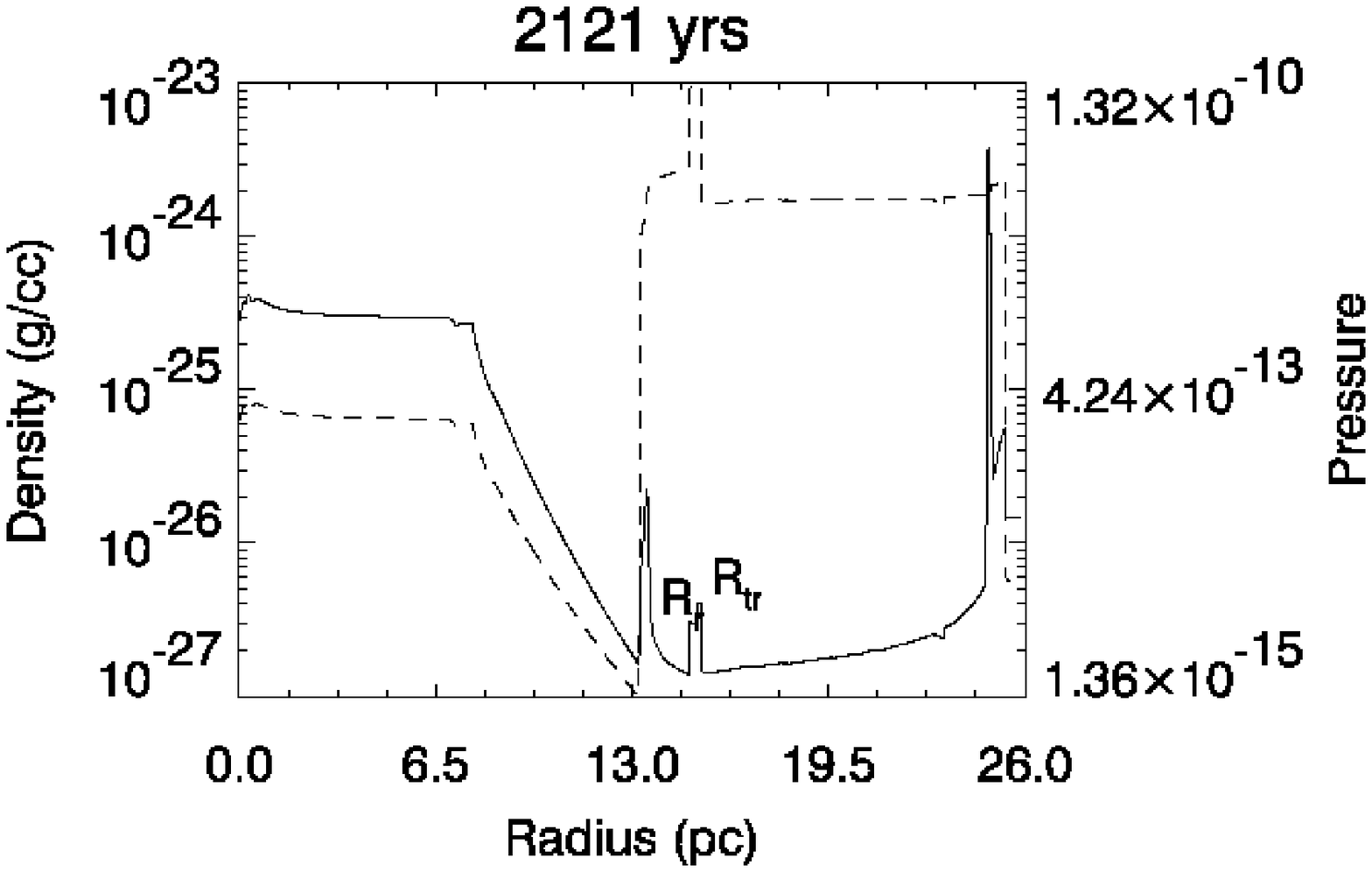}
\plottwo{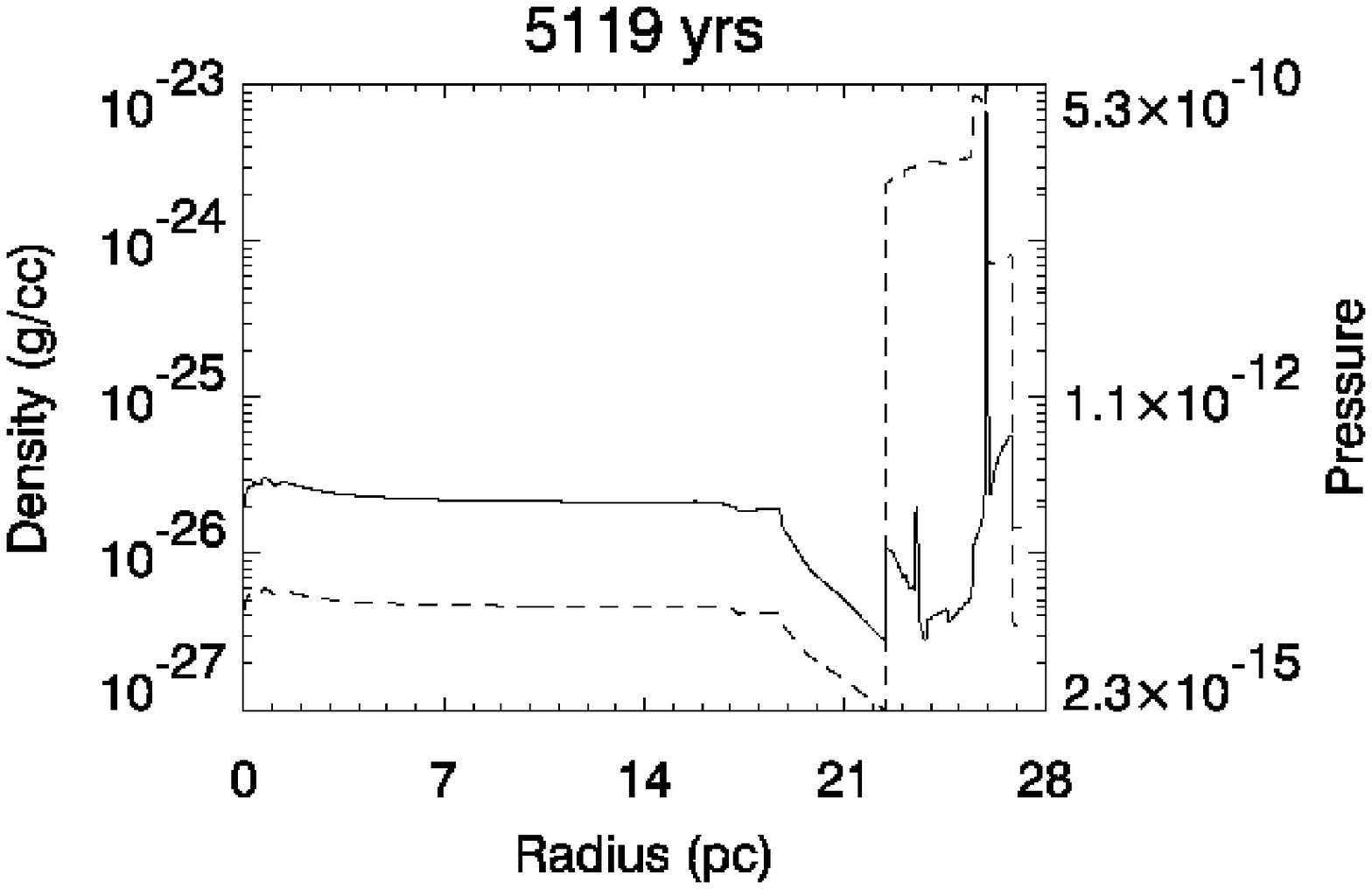}{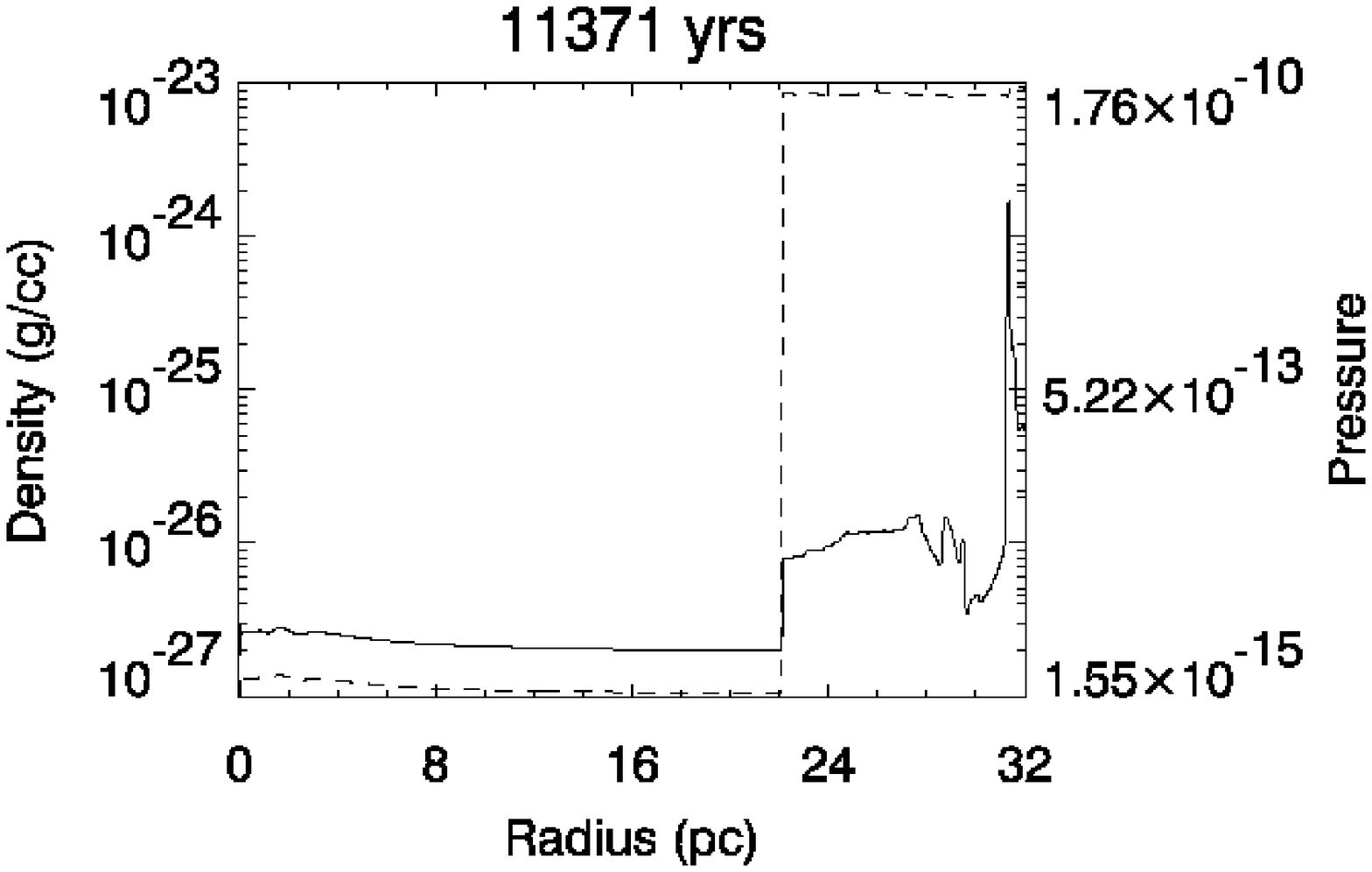}
\plottwo{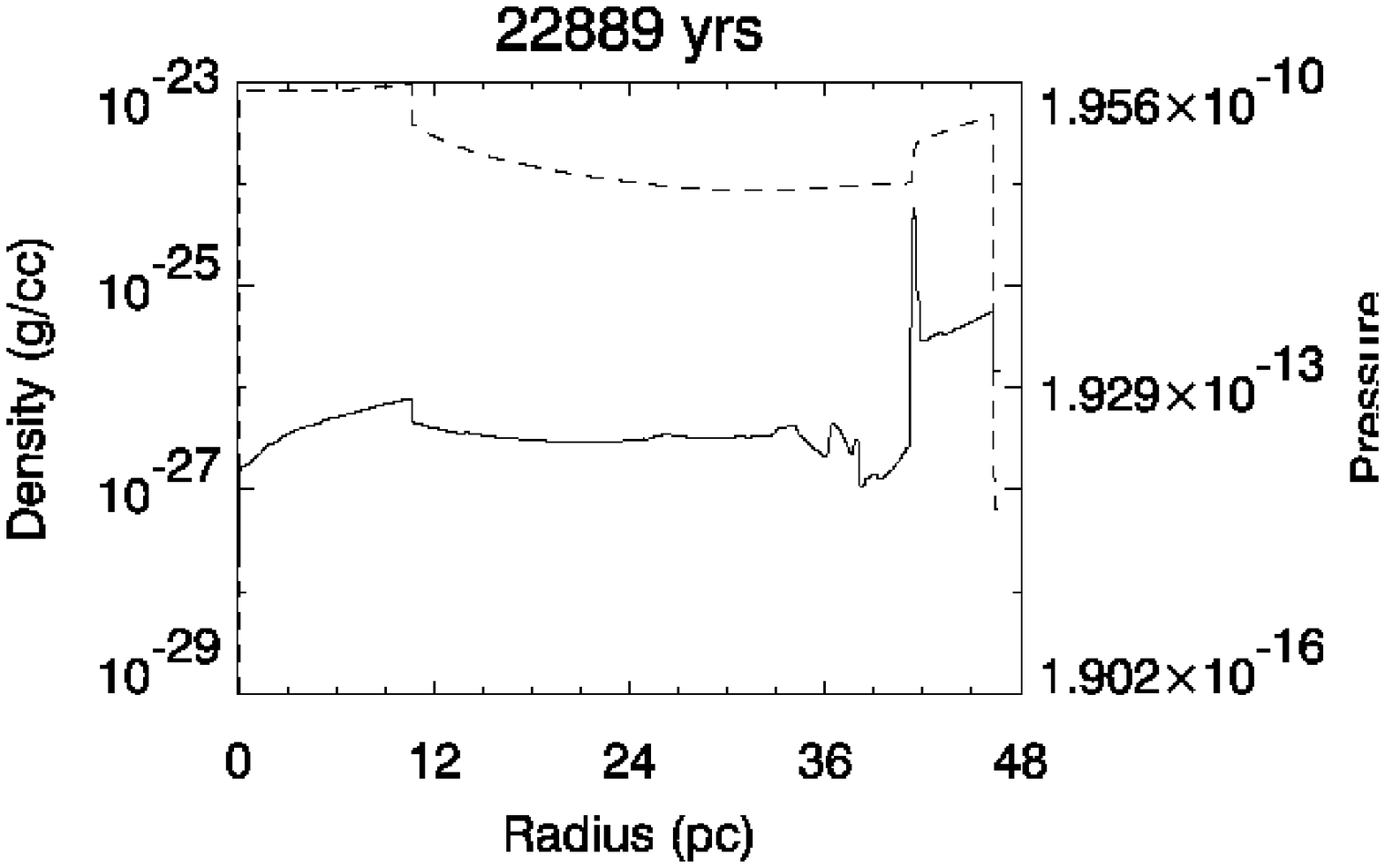}{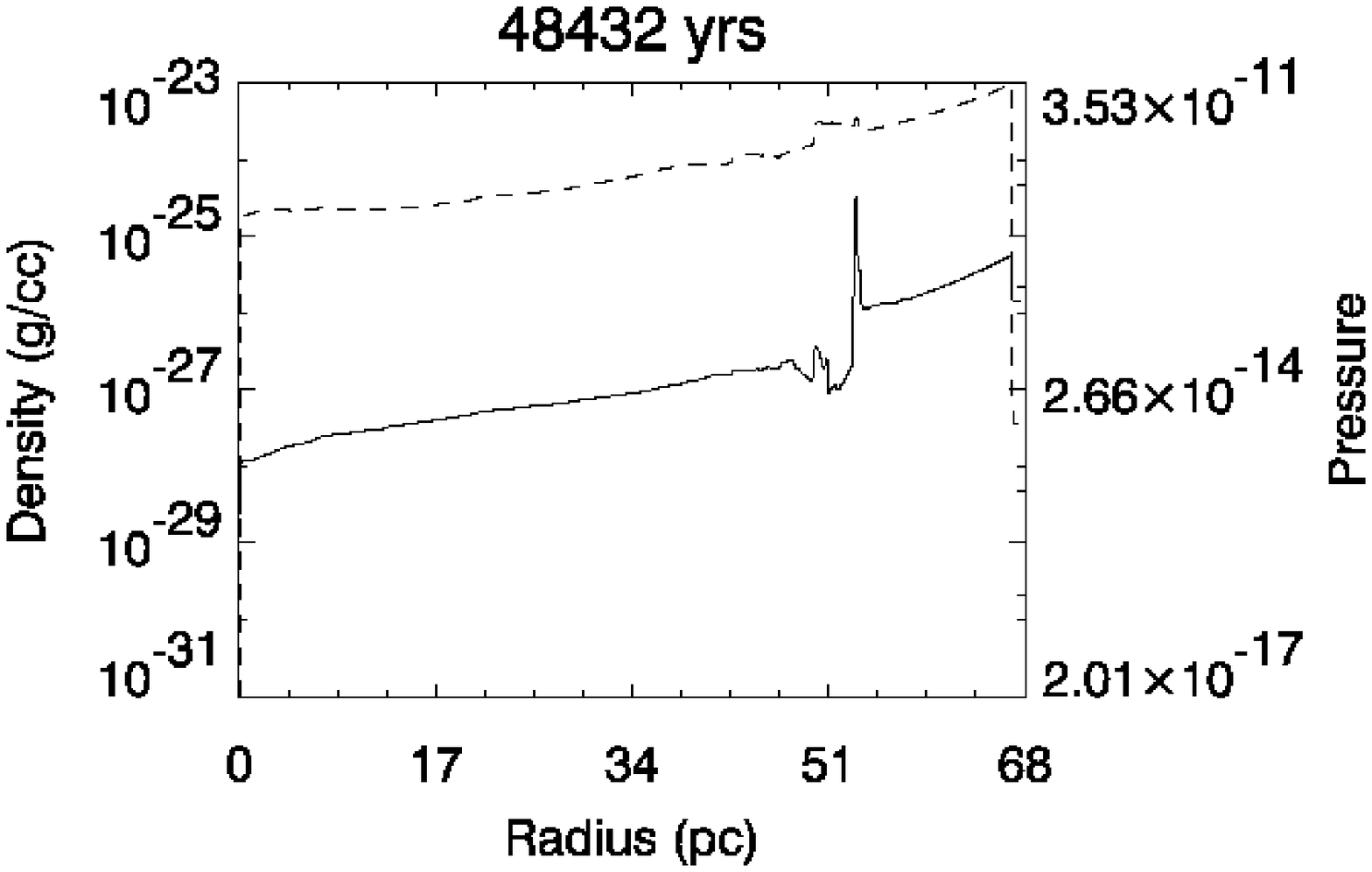}
\caption{Plots showing the density and pressure profile at various
times, for the simulation with $\Lambda = 3.7$ \label{fig:snbub037}}
\end{figure}

\begin{figure}
\includegraphics[angle=90, scale=0.6]{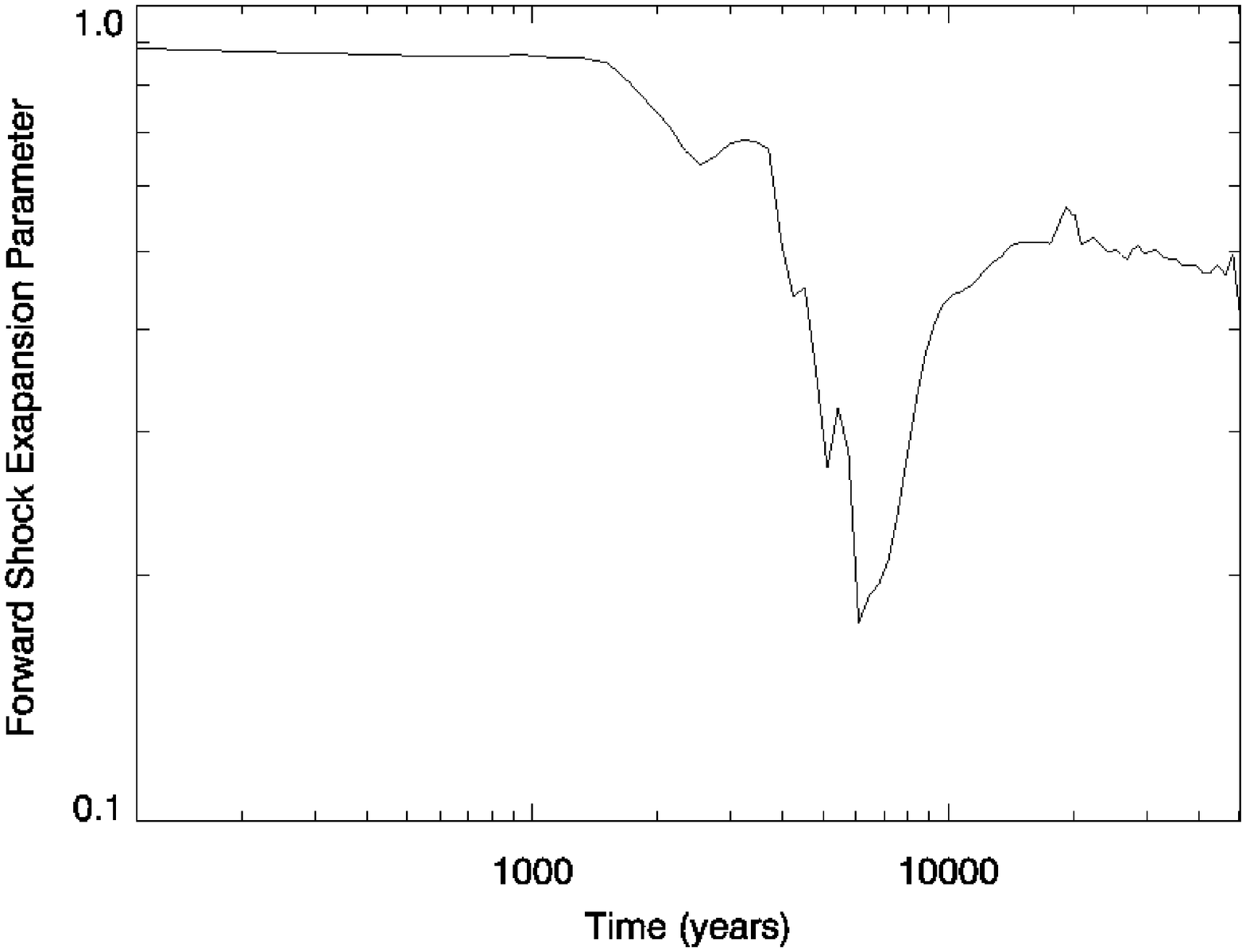}
\caption{The Evolution of the Expansion parameter for Case 2. \label{fig:exppar037}}
\end{figure}

Figure \ref{fig:snbub037} shows pressure and density snapshots at
various time intervals from a simulation of the evolution of the
remnant in this case. As seen from the plot at 112 years, the size of
the bubble is much larger than in Case 1. Note also the extremely low
density within a large section of the bubble interior, about 0.01
particles/cc. Such low density regions are characteristic of bubbles
around main sequence and Wolf-Rayet stars. Although the SN shock wave
takes about 2000 years to reach the wind termination shock, due to the
low wind density it has not swept up a very large mass of material
before that time, and therefore there is hardly any deceleration. The
expansion parameter (Fig \ref{fig:exppar037}) remains constant at a
value of 0.86 until the shock collides with the wind-termination
shock, giving rise to the usual double-shocked structure with a
reflected and transmitted shock. The transmitted shock then impacts
the wind-blown shell (frame 3, 5119 years) again resulting in a
reflected and transmitted shock. The impact also imparts considerable
momentum to the dense shell, causing the entire remnant to expand
outwards. The reflected shock attains high velocities, thermalizing
the inner ejecta as it moves towards the center. The transmitted
shock, which is expanding into a much higher density medium, moves
outwards with much lower velocities.

As expected from the discussion is \S \ref{sec:ts}, the forward shock
is expected to separate from the shell within a doubling time of the
shell. This behavior is illustrated in figure \ref{fig:snbub037},
where the shock has detached itself from the shell by the time it has
doubled its initial radius. The velocity profile resembles that for a
decelerating shock wave that is slowing down as it sweeps up more
material.

By about 20,000 years the reflected shock has converged onto the
center. A weak re-reflected shock is formed that expands outwards. The
impact of this shock with the dense shell does not result in an
appreciable increase in the X-ray luminosity.  Unlike Case 1, in this
case the reflected shock reaches the center before the transmitted
shock has separated itself significantly from the shell, and the
system does not easily ``forget'' the interaction.

It can be seen that although the remnant is almost 50,000 years old by
the end of the simulation, it does not yet display the profile
characteristic of the Sedov adiabatic stage. In particular, even after
the reflected shock has converged on the center and a weaker
re-reflected shock is present, the system is dominated by only the
forward shock wave, but the density profile is still different from
the Sedov solution. This illustrates an important characteristic of
SNRs evolving in low-density wind-blown bubbles - the evolution may
deviate from the classical description of ejecta-dominated, Sedov and
radiative stages.

When the reflected shock is moving towards the center, almost the
entire interior of the remnant is very hot, about 10$^8$ degrees. If
the density in the interior is high enough, the emission measure will
be large and the entire remnant will be visible in X-rays. In the
current example this is not the case, as the interior density is quite
low. But it is possible to envisage a situation where the shock-shell
interaction occurs earlier in the remnant's evolution (such as Case 1)
but the wind-blown shell still has mass larger than the ejecta. Then
the reflected shock will traverse much higher density ejecta on its
way to the center, and the emission from behind the reflected shock
could be comparable to that from the forward shock. The surface
brightness profile would then show X-rays all the way to the
center. It is possible that such an effect could be one explanation
for the so-called mixed-morphology remnants, which are X-ray bright
all the in way to the center.

\begin{figure}
\plotone{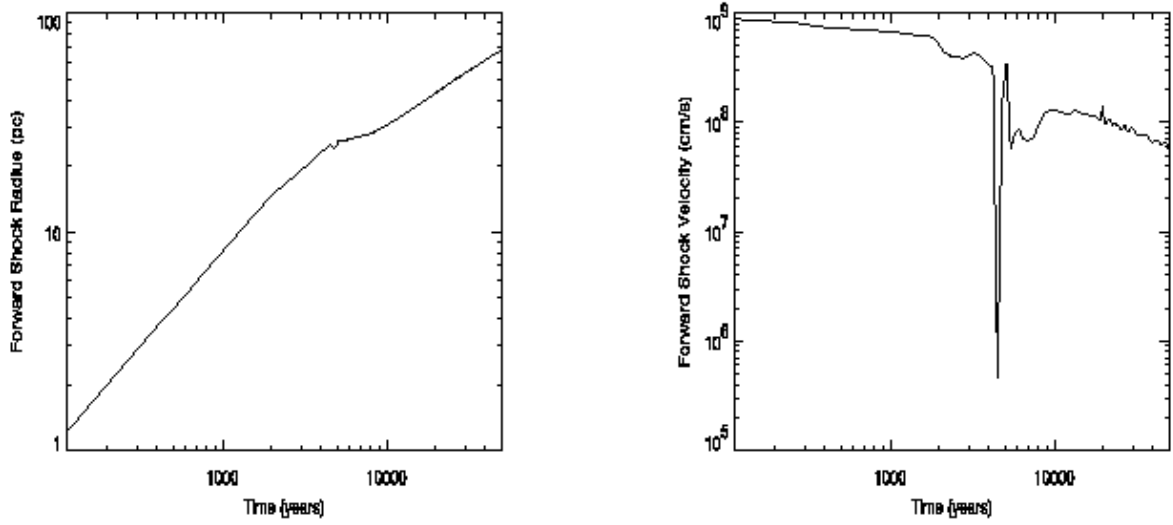}
\caption{Radius (left) and Velocity (right) plots for the evolution of
the SNR in Case 2. \label{fig:radvel037}}
\end{figure}

Figure \ref{fig:radvel037} shows the radius and velocity evolution of
the forward shock in this case, and Figure \ref{fig:exppar037} is a
plot of the expansion parameter evolution. In both cases they resemble
closely the counterparts in Case 1. The expansion parameter remains
constant up to the SN-wind termination shock interaction, then a
slight decrease until the shock-shell impact takes place, followed by
a significant decrease in the velocity and expansion parameter. As the
transmitted shock emerges the expansion parameter increases again, and
then gradually decreases till it approaches the Sedov value of 0.4 in
a constant density medium. It is clear that even though the behavior
of the reflected shock and the interior structure of the remnant is
quite different in both cases, this is not reflected in the behavior
of the outer shock. This serves as a cautionary note that studying the
behavior of the outer shock is not sufficient to provide information
regarding the evolutionary phase and structure of the remnant.

\begin{figure}
\includegraphics[angle=90, scale=0.7]{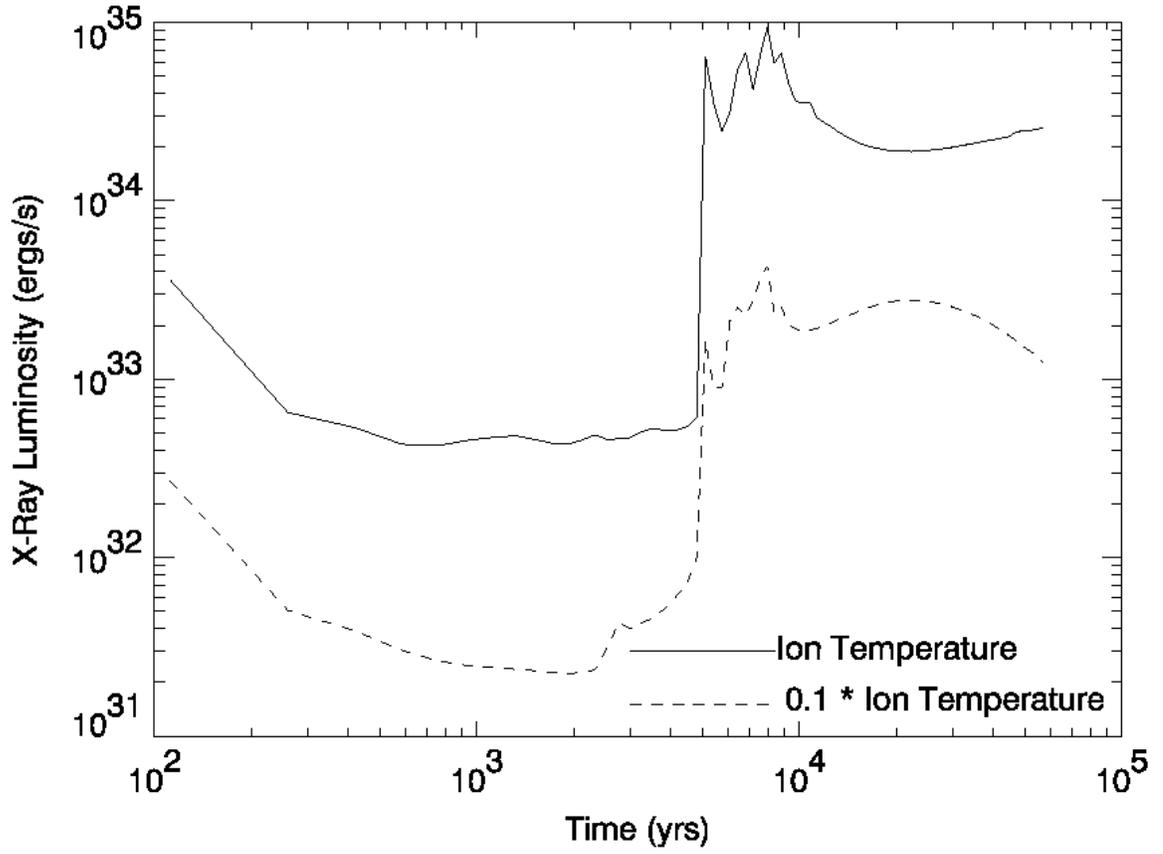}
\caption{The X-ray Luminosity During the Evolution of the remnant in
Case 2. For further details see Fig \ref{fig:xray014}.
\label{fig:xray037}}
\end{figure}

Figure \ref{fig:xray037} shows the X-ray luminosity of the remnant
over time. The overall evolution is similar to that in case 1,
although the increase in X-ray luminosity upon impact is smaller. In
this case the density is generally very small, and the temperature
high enough that line-emission is not important over almost the entire
evolution.  The X-ray luminosity is much lower in the beginning, given
the low density of the bubble interior. It is unlikely that such a
remnant could be observed by available X-ray instruments. The
luminosity jumps upon shock-shell impact. However the overall level is
a couple of orders of magnitude lower than in Case 1.

\begin{figure}
\includegraphics[scale=0.9]{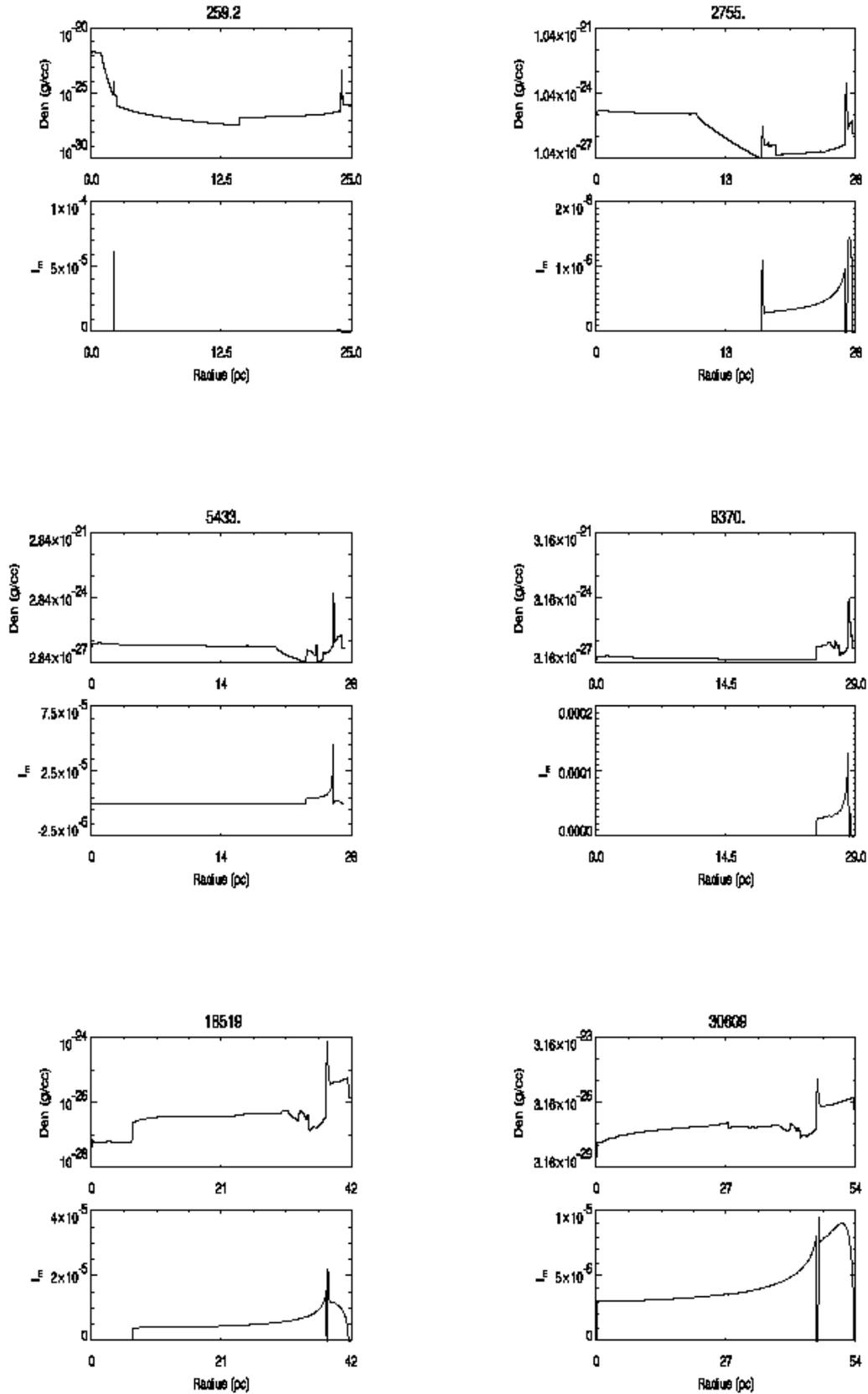}
\caption{X-ray surface brightness profiles during the Evolution of the
remnant in Case 2. \label{fig:xraysb037}}
\end{figure}

Fig \ref{fig:xraysb037} shows the approximate X-ray surface brightness
profiles, similar to Figure \ref{fig:xraysb014}. Again we start out
with the SN shock being the most visible source of X-rays. The
interaction of the SN shock wave with the wind-termination shock leads
to the expected rise in X-ray emission, and the interaction with the
wind-blown shell results in an even larger jump in the x-ray
intensity. Up to this point the X-ray evolution is similar to the
previous case. Beyond this however it departs considerably, because
the reflected shock does not have much time to interact with the
power-law ejecta before reaching the constant density inner
core. Therefore the remnant never takes on a double-shelled
appearance, but the major X-ray emitting region is always the outer
transmitted shock. As the reflected shock moves in the entire remnant
may be bright in X-rays. In the current case the emission measure is
low enough that the remnant may not be observable. But if the
shock-shell interaction happens early enough, then this is a
possibility worth investigating.

The reflected shock reaches the inner boundary before the transmitted
shock has advanced significantly. On reaching the center a
re-reflected shock is formed. This is expected in a realistic
situation when a spherical shock wave converges onto a compact
object. Note that the shock reflected off the boundary is quite weak,
and is not readily visible in X-rays (frame 6).

\subsection{$\Lambda >> 1$}
\label{sec:largel}

In some cases the shell mass may be considerably larger than the
ejecta mass. This can happen for instance if a massive star sheds a
large amount of mass during its evolution, leading to a pre-SN star
with a considerably smaller mass than the main-sequence value.  Stars
with initial mass greater than about 40 $\msun$ may end up with pre-SN
masses of 10 $\msun$ or less. In some cases it is possible that the
mass of the wind-blown shell is considerably larger than that of the
ejected material, and $\Lambda \ga 25$. In such extreme cases, when
the SN shock wave collides with the shell it will impart a
considerable fraction of the explosion energy to the shell. The shock
wave then becomes radiative, its velocity decreases considerably and
the time taken to emerge from the dense shell becomes extremely
large. In the limit that the shock loses most of its energy to the
shell, only a very weak transmitted shock will emerge from the
shell. The SN has then gone from the free-expansion phase to the
radiative phase, completely bypassing the Sedov phase. The reflected
shock will be quite strong and will thermalize the ejecta in a much
shorter time. The dense shell will expand with the additional energy
imparted to it, but the SN shock wave will not be visible as a
separate entity ahead of the dense shell. This case has been discussed
in detail by \citet{tbf90}.

Such cases, although not common, are quite plausible. Many WR nebulae
have masses that exceed 25-30 $\msun$. \citet{cec03} list some WR
nebulae with masses of hundreds to thousands of solar masses, mostly
made up of swept-up material. When the star explodes as a SN in these
nebulae the value of $\Lambda$ could be very large, possibly
50-100. In these cases the nebula will confine the SNR, and the
remnant will never emerge from the nebula. The size and dynamics of
the remnant are limited by that of the nebula, whose size is
essentially set by mass-loss in the main-sequence phase.

\section{Discussion}
\label{sec:disc}

It is clear that the medium around the progenitor star plays a large
role in shaping the further evolution of the remnant. Depending on the
nature of the progenitor star, the density of the medium into which
the SN shock expands may vary considerably. RSG progenitors will have
a region of very high density, surrounded by a low-density bubble
formed by the main-sequence star, bordered by a dense shell. WR
progenitors will have a low density wind-blown bubble surrounding the
star, bordered by a dense circumstellar shell. The interior of the
shell may show density variations depending on the various stages of
evolution prior to the SN stage.  As we have seen in the case of SN
1987A, BSG progenitors will form a cavity with density somewhat midway
between the above two cases. Factors such as ionization, rotation,
binary evolution and magnetic fields could all affect these results.

Quite often the SNR spends a not insignificant fraction of its life
within the bubble. It is then of interest whether the SN reaches a
Sedov stage while inside the bubble or not. Note that, depending on
the density profile, the swept-up mass must be several times that of
the ejecta mass before the remnant enters the Sedov stage \citep[see
for example][]{dc98}. Whether the remnant sweeps up enough material is
a function of the ejected mass and the bubble interior density. The
bubble density depends on the nature of the exploding star, which is a
function of its initial mass and mass-loss rate.  If the density into
which the SN shock evolves is very low, then the remnant will not
sweep up much material, and will remain in the free-expansion phase
for a much longer time than it would in the ISM. This is probably the
case with WR progenitors.  For remnants expanding in a low-density
bubble it is unlikely that they will reach the Sedov stage before they
collide with the dense shell. As mentioned in \S \ref{sec:largel} in
cases where the mass of the shell significantly exceeds that of the
ejecta, the remnant may evolve from the free-expansion stage directly
to the radiative stage, completely bypassing the Sedov stage.

If the progenitor is a RSG, then the surrounding density could be much
higher than that of the ISM. The remnant will sweep up a large mass of
material in a short time. Whether it evolves to the Sedov stage then
depends on the mass of the ejected material and the size of the RSG
wind-region, which depends primarily on the time spent in the RSG
stage. It is likely that the dense RSG wind will be surrounded by a
low density main-sequence bubble, whose density could be a few orders
of magnitude lower than that of the RSG wind. If the remnant has
already reached the Sedov stage then the evolution will continue as a
Sedov-type blast wave in a low density medium. If the remnant has not
reached the adiabatic stage then it is unlikely that it will do so in
the much lower density medium, and it will probably continue to be in
free-expansion until the shock wave collides with the circumstellar
shell.

The ejecta mass depends on the initial (zero-age main sequence) mass
of the star, the mass of the progenitor star just prior to the
explosion, and the nature of the compact stellar remnant. All of these
factors depend on the metallicity. According to recent work by
\citet{hfw03} and \citet{whw02}, stars in the mass range between 10
and 25 $\msun$ and solar metallicity will leave behind a neutron star,
and will have an ejecta mass about 10 solar masses or larger. However
much more massive stars can form larger mass black holes, and the
ejecta mass is reduced to just a few solar masses. A very broad
conclusion that can be drawn is that RSG stars will result in a large
ejecta mass, whereas WR stars will have a much lower ejecta mass (this
is not true for stars between 25 and 35 $\msun$). As we discussed
before SNe from WR progenitors will evolve in a much lower density
medium that those from RSG progenitors. Thus larger (smaller) ejecta
mass SNe will evolve in denser (low density) winds. A general
conclusion is that it will be difficult in most cases for the SNR to
sweep up enough material to reach the Sedov stage before colliding
with the dense shell.

The situation is opposite for SNe whose progenitor stars lie between
about 25 and 35 $\msun$, which will explode as RSGs but leave behind a
black hole with mass much larger than that of a neutron star, and a
very small mass of ejected material. These stars will most likely be
surrounded by a higher-density medium, and therefore will tend to
reach the Sedov stage quickly. When this Sedov remnant collides with
the dense shell, the reflected shock will traverse an almost evacuated
medium and move very quickly back to the center.

In higher mass stars the situation will be further complicated. The WR
stage is a highly evolved stage of a high-mass star greater than about
35 $\msun$. It is likely to be preceded by a RSG stage or perhaps even
a Luminous Blue Variable (LBV) stage, if the initial mass of the star
is much larger \citep{lhl94}. LBVs can have very large mass-loss
rates, and therefore very high-density winds, even compared to
RSGs. What this implies is that the WR stage will generally follow a
phase of very high-density winds. The WR phase itself has a high
mass-loss rate and wind velocity. The momentum of the WR wind is
generally going to be much larger than the RSG wind, and may even
exceed that of the LBV wind, although this depends on the specific
parameters. If the WR wind was preceded by a low-velocity RSG wind,
the WR wind momentum will carry the RSG wind material along with
it. This combined material may collide with the main sequence shell
and reflect back, forming more complicated structures than described
herein. The low density WR wind may end up containing higher-density
material from the RSG stage. The density profile will not be
representative of the wind-properties of the pre-SN progenitor star.

It is apparent that the medium into which a core-collapse SN evolves
cannot be easily categorized into either a constant density medium
such as the ISM, or a wind with a density $\rho \propto r^{-2}$. A
wind-blown bubble formed by two-wind interaction includes both
elements, a density that decreases as $r^{-2}$ followed by a more-or
less constant density region between the wind termination shock and
the contact discontinuity. However in stars larger than 35 $\msun$
mass-loss in the various evolutionary stages preceding the SN
explosion can give rise to very complicated density profiles beyond
the wind-termination shock. Close in to the star the density generally
follows a 1/r$^2$ law resulting from the last phase of evolution, but
further beyond the density may vary considerably depending on the
evolutionary sequence. These more realistic structures will be the
subject of future papers.

\section{Summary}
\label{sec:summary}

In this paper we have studied the evolution of supernova shock waves
in the circumstellar structures formed by pre-supernova mass loss. The
evolution depends mainly on one parameter $\Lambda$, the ratio of the
mass of the dense surrounding shell to the mass of the ejected
material. We have explored the changes in the evolution as the value
of this parameter increases, and found that the evolution is
significantly different from that for an explosion in the interstellar
medium. In particular as the value of $\Lambda$ increases we find that
the energy imparted to the shell increases, the evolution of the
remnant is speeded up, and the appearance of the remnant may change.

In contrast to the work of \citet[][hereafter TBFR90]{tbf90}, we have
used a much steeper density profile for the outer ejecta and a higher
ejecta mass. We have also explored more thoroughly the region around
$\Lambda \sim 1$, using two different values of $\Lambda=0.14$ and
$\Lambda=3.7$. Simulations were also carried out, although they are
not explicitly reported herein, with $\Lambda >> 1$. Our results are
consistent with the work of TBFR90 for $\Lambda >> 1$, as would be
expected. However, unlike TBFR90, we have explored two different cases
around the value $\Lambda \sim 1$ and find considerable difference
between the two cases. These differences can be noted in the evolution
as well as the X-ray appearance. In one case a double-shelled
structure is visible for a considerable amount of time after the shock
shell interaction. In the other case the double shell is mostly
absent, but if the interior density is high the entire remnant may be
visible in X-rays. However differences are not so easily apparent in
observations of the forward or outer shock front. Thus we caution that
inferring the properties simply by observing the outer shock
parameters can often lead to erroneous conclusions.

We have presented the evolution of the expansion parameter in each
case. It is worth pointing out again that the expansion parameter
varies significantly over the evolution, unlike in the case of
explosion within a constant density medium or a wind. This is very
important, because it implies that any instantaneous value of the
expansion parameter is not representative of the evolutionary stage of
the remnant, and will lead to an incorrect determination of
properties. And in particular, assuming that the remnant is in the
Sedov stage for computing the remnant parameters could lead to large
errors.

The initial evolution of the bubble, within a low-density medium, can
result in a large amount of time spent in the free-expansion phase, as
the swept-up mass remains quite small. The presence of the
circumstellar shell on the other hand, results in significant
deceleration of the SN shock wave, and decreases the time spent in the
Sedov stage as compared to the radiative stage. The net result is that
the Sedov stage could be significantly shortened, and in some cases
(see \S \ref{sec:largel}) may be entirely by-passed. It is plausible
however that the forward shock, after emerging from the CS shell, may
reach the Sedov stage. It is also quite likely though that given the
deceleration of the shock front, which we have seen will occur even
when the shell mass is about 10\% of the ejecta mass, that the remnant
may still reach the radiative stage much faster than in the absence of
the shell.

Our X-ray profiles show that for $\Lambda \la 1$, a double-shelled
structure is seen in X-rays, with emission coming mainly from shocked
shell and from behind the reverse shock. It is possible that such a
situation is present in Cas A. At later times when the outer shock has
separated significantly from the shell, the emission from the outer
shock begins to predominate. In the case with larger $\Lambda$ the
emission from the shocked shell always seems to predominate. It is
clear then that as the remnant evolves and the outer shock separates,
the X-ray emission from the shocked shell will appear to come from
deep within say the radio emission arising from behind the outer
shock. Precisely such a situation has been surmised for the remnant
W44 \citep{kh95}.

In this paper we have not considered multi-dimensional effects such as
hydrodynamical instabilities and turbulence. These aspects will be
dealt with in upcoming papers. Thin shells are known to be unstable,
due to many instabilities ranging from the well-known Rayleigh-Taylor
and Kelvin-Helmholtz type, to the range of instabilities lumped under
the heading of Vishniac-type instabilities.  These instabilities may
cause the shells to fragment, and lead to hydrodynamic behavior that
has not been explored herein. Furthermore, variations in pressure can
lead to turbulent behavior within the interior of the dense
shells. Also of significance is that, as the reflected shock in Case 1
moves back to the center, there is a region behind the reflected shock
that is Rayleigh-Taylor unstable. This instability is of importance
because it occurs in the interior of the remnant, far away from the
dense shell and forward shock, and can result in ejecta clumps and
filaments in the inner regions of the remnant.

Finally, the real test of these results will be to compare and
validate them with observed SNRs. We are indeed fortunate to be able
to closely witness the formation and evolution of a SN which occurred
within a wind-blown bubble, SN 1987A. The enormous body of
observational data that has been collected should enable us not only
to study this SN in the utmost detail, but by extension should provide
considerable information on the general theory of SN evolution in
wind-blown cavities. Of course SN 1987A has its own share of
complexities, and a thorough understanding of the circumstellar medium
requires 3-dimensional radiation hydrodynamics modeling, which has
not yet been carried out. Specific comparisons between our models and
observed remnants will be described in later papers in this series.

\acknowledgments Vikram Dwarkadas is supported by award \# AST-0319261
from the National Science Foundation. This work was initially started
as part of my PhD thesis, and it is a pleasure to acknowledge the
collaboration and guidance of Roger Chevalier at the University of
Virginia, who first encouraged me to work on this topic. I also thank
Roger for a careful reading of the paper. I am grateful to Chris McKee
for extremely stimulating discussions and helpful
suggestions. Suggestions from Bob Rosner, Kris Davidson and Roberta
Humphries are gratefully acknowledged.

\clearpage

%% Use the figure environment and \plotone or \plottwo to include
%% figures and captions in your electronic submission.
%% To embed the sample graphics in
%% the file, uncomment the \plotone, \plottwo, and
%% \includegraphics commands
%%
%% If you need a layout that cannot be achieved with \plotone or
%% \plottwo, you can invoke the graphicx package directly with the
%% \includegraphics command or use \plotfiddle. For more information,
%% please see the tutorial on "Using Electronic Art with AASTeX" in the
%% documentation section at the AASTeX Web site,
%% http://www.journals.uchicago.edu/AAS/AASTeX.
%%
%% The examples below also include sample markup for submission of
%% supplemental electronic materials. As always, be sure to check
%% the instructions to authors for the journal you are submitting to
%% for specific submissions guidelines as they vary from
%% journal to journal.

%% This example uses \plotone to include an EPS file scaled to
%% 80% of its natural size with \epsscale. Its caption
%% has been written to indicate that additional figure parts will be
%% available in the electronic journal.

\end{document}